\documentclass[aps,
twocolumn,
jcp,
floatfix]{revtex4-2}

\usepackage[english]{babel}
\usepackage{amsmath,amsthm,latexsym,amssymb,amsfonts,epsfig}
\usepackage{bm}
\usepackage{dsfont}
\usepackage[left = 1.35cm, right = 1.35cm, top = 1.98cm, bottom = 1.97cm]{geometry}

\usepackage{graphicx}
\usepackage{color}

\definecolor{CarmineRed}{rgb}{1.0, 0.0, 0.22}
\definecolor{OliveGreen}{rgb}{0.1, 0.4, 0.1}

\usepackage{setspace}
\emergencystretch=\maxdimen
\allowdisplaybreaks

\usepackage{mathtools}
\usepackage{mathrsfs}

\usepackage{type1ec} %

\usepackage{amsbsy}

\usepackage[colorlinks=true,citecolor=OliveGreen,linkcolor=OliveGreen,urlcolor=OliveGreen]{hyperref}

\newcommand{\ch}{\operatorname{ch}}
\newcommand{\sh}{\operatorname{sh}}
\renewcommand{\th}{\operatorname{th}} % this is redefined
\newcommand{\cth}{\operatorname{cth}}

\newcommand{\csh}{\operatorname{csh}} % csch = 1/sh
\newcommand{\sch}{\operatorname{sch}} % sech = 1/ch

\begin{document}

\title{Self-diffusiophoretic propulsion in wedge confinement: The role of phoretic interactions}

\author{Abdallah Daddi-Moussa-Ider}
\email{abdallah.daddi-moussa-ider@open.ac.uk}
\thanks{corresponding author.}
\affiliation{School of Mathematics and Statistics, The Open University, Walton Hall, Milton Keynes MK7 6AA, United Kingdom}

\author{Ramin Golestanian}
\email{ramin.golestanian@ds.mpg.de}
\affiliation{Max Planck Institute for Dynamics and Self-Organization (MPI-DS), 37077 Göttingen, Germany}
\affiliation{Rudolf Peierls Centre for Theoretical Physics, University of Oxford, Oxford OX1 3PU, United Kingdom}

\begin{abstract}
We investigate the self-diffusiophoretic motion of a catalytically active spherical particle confined within a wedge-shaped domain. Using the Fourier-Kontorovich-Lebedev transform, we solve the Laplace equation for the concentration field in the diffusion-dominated regime. The method of images is employed to obtain the first and second reflections of the concentration field, accounting for both monopole and dipole contributions of the particle’s surface activity. Based on these results, we derive leading-order expressions for the self-induced phoretic velocity in the far-field limit and examine how it varies with the wedge opening angle and the particle’s position within the domain. We focus on the contributions to the phoretic velocities arising from phoretic interactions, without accounting for hydrodynamic effects. Our findings reveal that the wedge geometry significantly affects both the magnitude and direction of particle motion. Our study provides a systematic framework for calculating the contributions to the phoretic velocity arising from concentration disturbances near corners, with implications for microfluidic design and control of autophoretic particles in confined geometries.

\end{abstract}

\maketitle

\section{Introduction}

Active matter has emerged as a vibrant area of research, attracting increasing interest in the biophysics and bioengineering communities~\cite{Chen2025,lauga09, zottl16, bechinger16, needleman2017active, zottl2023modeling, te2025metareview,Gompper2020}. Of particular focus are self-propelling microswimmers, which provide model systems for probing the principles of out-of-equilibrium dynamics in biological and cellular contexts. Active particles are capable of autonomous motion, converting energy harvested from their environment into mechanical work. Beyond their fundamental significance, such systems offer promising avenues for biomedical applications, including targeted drug delivery, precision nanosurgery and diagnostic imaging~\cite{Ju2025,park2017multifunctional, tang2020enzyme, soto2020medical, llacer2021biodegradable}. In addition to their individual motion, collections of active particles exhibit rich collective behaviors, leading to a variety of complex phenomena, including dynamic clustering, swarming, pattern formation, and emergent transport properties that are absent in passive systems~\cite{golestanian2012collective,theurkauff2012dynamic, pohl2014dynamic, saha2014clusters, soto2015self, liebchen2018synthetic}.

Phoretic self-propulsion has become a prominent mechanism in active matter research~\cite{golestanian_les_houches,moran17, illien17}.  Self-phoretic swimmers achieve autonomous motion by harnessing local interactions with the surrounding medium. The motion of phoretic active colloids is driven by effective slip velocities arising from local concentration gradients produced by chemical reactions occurring at their surface~\cite{golestanian05,golestanian07, michelin14}. 
However, phoretic motion is not solely driven by an effective slip velocity; it can also arise from body-force-dipole distributions that extend into the surrounding fluid rather than being confined to a thin layer~\cite{sabass2012nonlinear, sharifi13,AgudoCanalejo2018}.
Numerous studies have explored the influence of geometric confinement on the dynamics and behavior of individual self-phoretic particles~\cite{uspal15, ibrahim15, ibrahim2016walls, das2015boundaries, simmchen16,uspal16,  mozaffari16, bayati2019dynamics, popescu2018effective}. 
These studies show that boundaries, walls, and nearby surfaces can profoundly affect particle propulsion, orientation, and trajectories. For example, planar walls can induce solute gradients which, in combination with hydrodynamic interactions, can enhance motion along the wall, reorient the particle away from the surface, or in some cases lead to a hovering state in which the particle remains stationary. Similarly, active particles near shear flows and bounding surfaces can exhibit complex behaviors, such as rheotaxis, even for symmetric spherical swimmers~\cite{sharan2022upstream}.
Beyond single-particle dynamics, the collective behavior of multiple phoretic colloids has been investigated in diverse contexts, revealing complex interactions and emergent phenomena~\cite{varma2019modeling, sharan2023pair, OuazanReboul2023b}.

Breaking spatial symmetry is a key requirement for phoretic self-propulsion at low Reynolds numbers. Geometric asymmetries can be exploited to generate chemical gradients, enabling self-propulsion even in isotropic self-phoretic particles.
Through phoretic and hydrodynamic interactions with inert, non-motile particles, isotropic active particles have been shown to be capable of swimming by forming dynamic clusters~\cite{soto2014self,michelin2015autophoretic, varma2018clustering, picella2022confined, nasouri2020exact}. Moreover, internal phoretic flows may arise solely from geometric asymmetries on chemically homogeneous surfaces~\cite{michelin2015geometric, lisicki2016phoretic,nowak2025diffusioosmotic}.

Here, we employ a far-field approach to study the diffusiophoretic motion of a catalytically active spherical colloid confined within a wedge-shaped domain, bounded by a single straight, infinitely extended edge that forms the apex of the wedge.
While diffusiophoresis near a no-slip wall~\cite{uspal15, ibrahim15, ibrahim2016walls, das2015boundaries, simmchen16,uspal16,  mozaffari16, bayati2019dynamics, popescu2018effective} or a fluid–fluid interface~\cite{malgaretti2018self, daddi2022diffusiophoretic} has been investigated extensively in earlier studies, a systematic analysis of the behavior of phoretically active colloids in the vicinity of a corner remains largely unexplored.
In many biologically relevant applications, wedges are ubiquitous geometrical features that arise in confined environments, such as at cell junctions and within microstructured biological media.
For instance, two-dimensional computer simulations have shown that a static chevron-shaped wall represents an efficient trap for the collective capture of self-propelled colloidal rods~\cite{kaiser2012capture}.
The present contribution constitutes an initial step toward characterizing the dynamics of active particles under wedge confinement.

In the context of hydrodynamics, flows near three-dimensional corners were first investigated through the analysis of a sequence of increasingly complex wedge-shaped geometries~\cite{sano76, sano1977slow, sano1978effect, hasimoto80, sano1977slow_thesis}. Three-dimensional Stokes flow in the vicinity of a corner has been revisited using methods from complex analysis~\cite{dauparas2018leading, dauparas2018stokes}. Later, the dynamics of a microswimmer confined by a wedge-shaped free-slip interface has been explored using the method of images for different wedge opening angles~\cite{sprenger2023microswimming}. More recently, Green’s functions for a linearly elastic, homogeneous, and isotropic material in a wedge-shaped geometry have been derived for various types of boundary conditions~\cite{daddi2025proc, daddi2025jelasticity}. Under the assumption of incompressibility, the resulting expressions are also applicable to low-Reynolds-number viscous hydrodynamic flows.

In this work, we employ the Fourier-Kontorovich-Lebedev transform to address the diffusiophoresis problem near a wedge, a technique well suited for solving boundary-value problems in geometries with radial symmetry and angular confinement, such as wedges and cones. The method of images is applied to compute the concentration field up to the second reflection, accounting for both monopole and dipole contributions. From these results, we determine the self-induced diffusiophoretic velocities and analyze their dependence on the system geometry, specifically the wedge opening angle and the particle’s position within the domain.
In this article, we focus exclusively on the contributions to the phoretic velocities arising from phoretic interactions and do not consider the effects of hydrodynamic interactions.
The latter requires the knowledge of higher-order hydrodynamic singularities of Stokes flow in a wedge geometry, which are not yet available. For this reason, a full treatment of the hydrodynamic interactions near a wedge is left for future work, once these higher-order singularities have been determined.

The paper is structured as follows. In Sec.~\ref{sec:math}, we introduce the problem setup and specify the boundary conditions, and present the Fourier-Kontorovich-Lebedev transform for solving the Laplace equation in the diffusion-dominated regime. Sec.~\ref{sec:concentration} presents the solutions for the first and second reflections of the concentration field, considering the first two moments of the particle’s surface activity, along with contour plots illustrating the concentration distribution. In Sec.~\ref{sec:phoretic_velocity}, we derive leading-order expressions for the self-induced phoretic velocity in the far-field limit, when the active particle is distant from the nearest wedge surface, and discuss its variation with system parameters. Sec.~\ref{sec:planar_boundary} demonstrates how the planar boundary solution is recovered for a wedge opening angle of~180$^\circ$. 
Sec.~\ref{sec:superposition} presents results on the use of the superposition approximation to predict the induced phoretic velocity.
Finally, we summarize our findings and conclude in Sec.~\ref{sec:conclusions}.

\section{Mathematical model}
\label{sec:math}

\begin{figure}
    \centering
    \includegraphics[width=\linewidth]{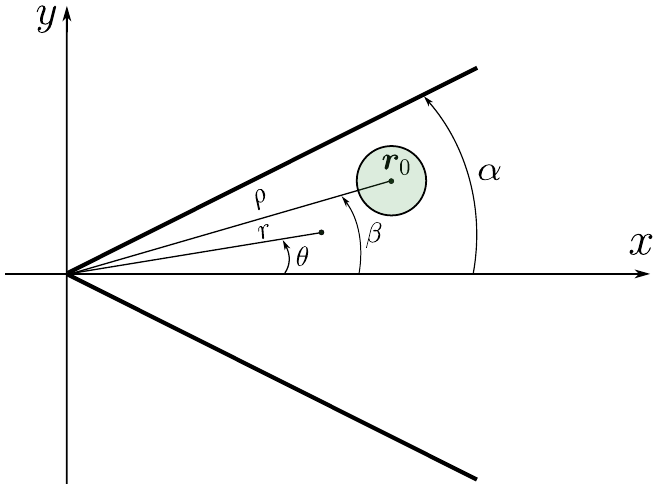}
    \caption{A catalytically active particle of radius $R$ is located near a wedge formed by two walls intersecting along the $z$-axis, with the wedge having a semi-opening angle $\alpha$. The evaluation point is denoted by $(r,\theta,z)$ in cylindrical coordinates, while the position of the active colloid is specified by the polar distance~$\rho$ and polar angle~$\beta$. No-flux boundary conditions are assumed at the surfaces of the walls.
    }
    \label{fig:illustration}
\end{figure}

We study the motion of a self-propelling, catalytically active colloidal particle of spherical shape of radius $R$ in a viscous fluid confined within a wedge-shaped domain with a straight edge forming its tip. The system geometry is shown in Fig.~\ref{fig:illustration}. A cylindrical coordinate system $(r, \theta, z)$ is used, with the wedge’s straight edge aligned along the $z$-axis. The fluid is confined by the wedge surfaces at $\theta = \pm\alpha$, where $\alpha \in (0, \pi)$; the case $\alpha = \pi/2$ corresponds to a semi-infinite fluid bounded by a flat surface. The particle is located at $\bm{r}_0$, corresponding to $(r, \theta, z) = (\rho, \beta, 0)$.
We denote by $\hat{\bm{u}}$ the orientation vector pointing along the axis that defines the direction of self-propulsion, which is determined by the particle’s surface activity, assumed to be axisymmetric. For example, a Janus particle, half coated with a catalytic material, produces solute on only one side, the orientation vector points from the inert side toward the active (catalytic) side. Here, $\hat{\bm{u}}$ is a three-dimensional vector representing any possible orientation of the particle.
No-flux boundary conditions are applied at the wedge surfaces.

Just to clarify terminology, we define an \textit{obtuse wedge} as one with a semi-opening angle $\alpha \in (0, \pi/2)$, and a \textit{salient wedge} as one with $\alpha \in (\pi/2, \pi)$.
In contrast to many elasticity or hydrodynamics theories that are often limited to obtuse wedges~\cite{sano1978effect, daddi2025proc, daddi2025jelasticity}, here $\alpha$ can take any value between 0~and~$\pi$, allowing exploration of the entire parameter space.

We assume that the fluid dynamics are governed by low-Reynolds-number hydrodynamics, where inertial effects are negligible compared to viscous effects~\cite{happel12,kim13}. Additionally, we consider the zero-P\'{e}clet-number limit, so that the concentration field around the particle is governed solely by diffusion, and the particle’s motion does not significantly advect the surrounding solute.
For microscale active colloids moving in typical solutions, these modeling assumptions are appropriate.

\subsection{Governing equations for phoretic particles}

Exploiting the geometry of the system, we determine the concentration field using cylindrical coordinates.
The solute concentration is governed by the Laplace equation
\begin{equation}
    \boldsymbol{\nabla}^2 c = 0 \, ,
    \label{eq:Laplace}
\end{equation}
which, in cylindrical coordinates, is written as
\begin{equation}
    \frac{1}{r} \frac{\partial}{\partial r} \left( r \,  \frac{\partial c}{\partial r} \right) + \frac{1}{r^2} \frac{\partial^2 c}{\partial \theta^2} + \frac{\partial^2 c}{\partial z^2} = 0 \, .
\end{equation}

We define the position vector relative to the center of the active colloid as $\bm{s} = \bm{r} - \bm{r}_0$, with $s = |\bm{s}|$ representing the distance between the evaluation point and the center of the particle. We further define the unit vector $\hat{\bm{s}} = \bm{s}/s$.

We prescribe the boundary condition at the surface of the active colloid as
\begin{equation}
    \left. -D \, \hat{\bm{n}} \cdot \boldsymbol{\nabla} c (\hat{\bm{s}}) \, \right|_\mathcal{S}
    = A (\hat{\bm{s}}) \, , 
    \label{eq:BCs_sphere}
\end{equation}
with $\mathcal{S}$ representing the particle's surface and $\hat{\bm{n}}$ its unit normal vector.
Note that $\hat{\bm{n}} \equiv \hat{\bm{s}}$ for the case of a sphere, as considered here, but for non-spherical particles, $\hat{\bm{n}}$ and~$\hat{\bm{s}}$ generally differ: $\hat{\bm{n}}$~remains the local outward normal to the surface, while~$\hat{\bm{s}}$~points from the particle center to a surface point and does not necessarily align with the normal direction.
In addition, $A$ represents the surface activity, quantified by the radial flux at the colloid surface, modeling the catalytic reaction on the active cap.
Here, we prescribe the surface activity in terms of Legendre polynomials as
\begin{equation}
    A (\hat{\bm{s}}) = \sum_{\ell = 0}^\infty A_\ell P_\ell \left( \hat{\bm{u}} \cdot \hat{\bm{s}} \right) ,
\end{equation}
where $A_\ell$ represent the surface activity moments.

We assume no-flux boundary conditions at the wedge walls located at $\theta = \pm\alpha$, expressed as
\begin{equation}
    \left. \, \hat{\bm{e}}_\theta \cdot \boldsymbol{\nabla} c \, \right|_{\theta=\pm\alpha} 
    = \left. \frac{\partial c}{\partial \theta} \right|_{\theta=\pm\alpha}
    = 0 \, .
    \label{eq:BCs_wedge}
\end{equation}

As a regularity condition, the solute concentration approaches the constant value $c_\infty$ far from the active colloid. 

The procedure for determining the dynamics of an active colloid is well established in the field of phoretic active matter~\cite{golestanian_les_houches}.
In this context, particle motion arises from self-generated concentration gradients. The resulting slip velocity is determined by the tangential gradient of the solute concentration through the phoretic mobility~$\mu$. 
Specifically~\cite{golestanian05, golestanian07},
\begin{equation}
    \bm{v}_\mathrm{S} = \mu 
    \left( \bm{I} - \hat{\bm{n}} \hat{\bm{n}} \right) \cdot \boldsymbol{\nabla} c \, .
    \label{eq:vs}
\end{equation}

To circumvent the need to explicitly solve for the hydrodynamic fields, the reciprocal theorem of fluid mechanics provides a powerful and efficient framework~\cite{masoud2019reciprocal}. This approach relies on using the solution of a suitably chosen auxiliary problem to obtain the velocities in the primary problem of interest. In the present case, the auxiliary problem is taken to be that of a passive particle subjected to externally applied hydrodynamic force and torque~\cite{stone96}. By exploiting the no-slip condition on the wedge surface and the fact that the active particle is force and torque free, we obtain the following result \cite{uspal15}
\begin{equation}
    \bm{V} \cdot \bm{F}' + \boldsymbol{\Omega} \cdot \bm{L}'
    = -\oint_\mathcal{S} \bm{v}_\mathrm{S} \cdot \boldsymbol{\sigma}' \cdot \hat{\bm{n}} \, \mathrm{d}S \, .
    \label{eq:RT}
\end{equation}
Here
\begin{subequations}
    \begin{align}
    \bm{F}' &= \oint_\mathcal{S} \boldsymbol{\sigma}' \cdot \hat{\bm{n}} \, \mathrm{d}S \, , \\
    \bm{L}' &= \oint_\mathcal{S}
    \left( \bm{r}-\bm{r}_0 \right) \times
    \boldsymbol{\sigma}' \cdot \hat{\bm{n}} \, \mathrm{d}S \, , 
\end{align}
\end{subequations}
represent the hydrodynamic force and torque, respectively, as exerted by the quiescent fluid on the particle undergoing steady-state translation or rotation, subject to a no-slip boundary condition on its surface.
We note that $\boldsymbol{\sigma}'$~represents the viscous stress tensor in the auxiliary problem.

We introduce the dimensionless parameter $\epsilon = R/d$, where $d = \rho \sin(\alpha-\beta)$ denotes the distance between the center of the sphere and its orthogonal projection onto the nearest wall. The present analysis is carried out in the far-field limit, focusing on the regime where $\epsilon \ll 1$.

In the presence of a confining wedge, the leading-order correction to $\boldsymbol{\sigma}'$ is expected to scale as $F/d^2$ for an applied force and as $L/d^3$ for an applied torque, based on simple dimensional arguments. This scaling can be explicitly verified for a planar wall using the Blake tensor~\cite{blake1971note} Consequently, the translational and rotational velocities of the particle can be obtained using the corresponding expressions for a particle in an unbounded fluid \cite{ibrahim2016walls}
\begin{equation}
    \bm{V} = -\langle \bm{v}_\mathrm{S} \rangle \, , \quad
    \boldsymbol{\Omega} = -\frac{3}{2R} \,
    \langle \hat{\bm{n}} \times \bm{v}_\mathrm{S} \rangle \, ,
    \label{eq:transl_rot}
\end{equation}
where $\langle f \rangle$ denotes the average of a quantity $f$ over the surface of the active colloid, defined as 
\begin{equation}
    \langle f \rangle = 
    \frac{1}{4\pi R^2} \oint_\mathcal{S} f \, \mathrm{d}S \, .
\end{equation}

We show that the monopole and dipole contributions give rise to translational phoretic velocities scaling as $\epsilon^2$ and~$\epsilon^3$, respectively. Accordingly, including corrections in $\boldsymbol{\sigma}'$ in Eq.~\eqref{eq:RT} would introduce higher-order effects beyond the scope of the present study. Specifically, such corrections would generate terms scaling as $\epsilon^4$ for the monopole contribution and $\epsilon^5$ for the dipole contribution, and are therefore justifiably neglected.

\subsection{Fourier-Kontorovich-Lebedev transform}

To determine the concentration field, we employ the well-established Fourier-Kontorovich–Lebedev transform, which is particularly suited for boundary value problems in wedge-shaped geometries. This approach transforms the axial and radial coordinates into the corresponding axial and radial wavenumbers, denoted by~$k$ and~$\nu$, respectively.
While the Fourier transform is widely known and its use is well understood, the Kontorovich–Lebedev transform is far less familiar. It was originally introduced by the Russian mathematicians Kontorovich and Lebedev to address specific classes of boundary value problems~\cite{kontorovich1938one, kontorovich1939method}. Its mathematical foundations and further applications were subsequently developed~\cite{kontorovich1939application, lebedev1946, lebedev1949}; see Erd{\'e}lyi \textit{et al.}~\cite[p.~75]{erdelyi1953higher} for additional details. The Kontorovich–Lebedev transform has since been applied to a variety of physical problems involving wedge geometries, most notably in electromagnetic scattering and diffraction~\cite{lowndes1959application, rawlins1999diffraction, antipov2002diffraction, salem2006electromagnetic}, elasticity~\cite{daddi2025proc, daddi2025jelasticity}, as well as in studies of fluid flows~\cite{waechter1969steady, waechter1969steadyB}.

We begin by defining the forward Fourier transform of the concentration field $c(r,z)$ with respect to $z$ as
\begin{equation}
    \hat{c}(r, \theta, k) 
    := \mathscr{F} \left\{ c \right\}
    = 
    \int_{-\infty}^\infty c(r,\theta,z) \, e^{ikz} \, \mathrm{d} z \, , 
\end{equation}
Next, we introduce the forward Kontorovich–Lebedev transform with respect to $r$ as
\begin{equation}
    \hspace{-0.1cm}
    \widetilde{c}(\nu, \theta, k) 
    := \mathscr{K}_{i\nu} \left\{ \hat{c} \right\}
    = \int_0^\infty \hat{c}(r, \theta, k) K_{i\nu} (|k|r) \, r^{-1} \, \mathrm{d} r  \, ,
\end{equation}
where $K_{i\nu}(|k|r)$ is the modified Bessel function of the second kind~\cite{abramowitz72} of imaginary order $i\nu$. Note that the polar angle $\theta$ is not affected by these transformations. 

Hereafter, we denote the hyperbolic sine, cosine, tangent, and cotangent functions by sh, ch, th, and cth respectively. 
In addition, we denote by sch and csh the hyperbolic secant and hyperbolic cosecant, respectively.

The inverse transforms are then expressed as
\begin{equation}
    \hat{c}(r,\theta,k) = 
    \frac{2}{\pi^2} \int_0^\infty \widetilde{c}(\nu, \theta, k) K_{i\nu} (|k|r) \sh (\pi\nu) \, \nu \, \mathrm{d} \nu \,  \label{eq:inv_KL}
\end{equation} 
for the inverse Kontorovich-Lebedev transform, and
\begin{equation}
    c(r,\theta,z) = \frac{1}{2\pi} 
    \int_{-\infty}^\infty \hat{c}(r,\theta, k) \, e^{-ikz} \, \mathrm{d} k \,  \label{eq:inv_F}
\end{equation}
for the inverse Fourier transform.
In many practical cases, performing the inverse Fourier transform is considerably simpler. Consequently, the solution is often expressed as a single infinite integral over the radial wavenumber~$\nu$.

We define the notation for the combined transform as
\begin{equation}
    \widetilde{c} = \mathscr{T}_{i\nu} \left\{ c \right\}
    = \mathscr{K}_{i\nu} \left\{ \mathscr{F} \left\{ c \right\} \right\} \, ,
\end{equation}
which we denote as the FKL transform.
As a result, in FKL space, the Laplace equation, Eq.~\eqref{eq:Laplace}, governing the concentration field, is reduced to a second-order ordinary differential equation in the polar angle $\theta$, of the form~\cite{daddi2025proc}
\begin{equation}
     \left( \frac{\partial^2}{\partial\theta^2}-\nu^2 \right) \widetilde{c} = 0 \, .
     \label{eq:Laplace_FKL}
\end{equation}
A general solution to Eq.~\eqref{eq:Laplace_FKL} is given by
\begin{equation}
     \widetilde{c} = K_1 \sh (\theta\nu) + K_2 \ch (\theta\nu) \, ,
     \label{eq:solution_form}
\end{equation}
where $K_1$ and $K_2$ are functions of $k$ and $\nu$, to be determined from the boundary conditions.

In the following, we summarize well-known results from the literature on the concentration field and the resulting self-diffusiophoretic motion of a colloidal particle in a bulk fluid. We then address the current problem of determining the concentration field and phoretic velocities in the presence of the wedge.

\section{Concentration field}
\label{sec:concentration}

\subsection{Motion in a bulk medium}

The general solution of the Laplace equation~\eqref{eq:Laplace} in spherical coordinates can be written as a series expansion in spherical harmonics.
Since we are dealing with an axisymmetric problem, the solution can be expressed using Legendre polynomials, involving only the exterior harmonics.
In a bulk fluid medium, i.e., in the absence of the confining wedge, the concentration field follows from the flux boundary condition prescribed by Eq.~\eqref{eq:BCs_sphere} as 
\begin{equation}
    c^{(0)} = c_\infty + \sum_{\ell = 0}^\infty
    q_\ell \left( \frac{R}{s} \right)^{\ell+1} 
    P_\ell \left( \hat{\bm{u}} \cdot \hat{\bm{s}} \right) ,
\end{equation}
where we have defined the scaled activity moments as
\begin{equation}
    q_\ell = \frac{A_\ell}{\ell+1} \frac{R}{D} \, .
    \label{eq:q_ell}
\end{equation}

We note that the moments of the surface activity are readily obtained by exploiting the orthogonality of the Legendre polynomials as
\begin{equation}
    A_\ell = \left( \ell+\frac{1}{2} \right) \int_{-1}^1 A\left( \hat{\bm{s}} \right) P_\ell \left( \hat{\bm{u}} \cdot \hat{\bm{s}} \right) \, \mathrm{d} \left( \hat{\bm{u}} \cdot \hat{\bm{s}} \right) \, .
\end{equation}

In the present contribution, we focus on the first two activity moments, i.e., we set $A_\ell = 0$ for $\ell \ge 2$.
The influence of higher-order moments can be analyzed using the same approach outlined below.
Accordingly, the solute concentration in an unbounded medium can be expressed as
\begin{equation}
    c^{(0)} = c_\infty + q_0 \, \frac{R}{s} 
    + q_1 \left( \frac{R}{s} \right)^2 
    \hat{\bm{u}} \cdot \hat{\bm{s}} \, .
    \label{eq:c0}
\end{equation}
The second and third terms correspond to the source monopole and source dipole singularities, respectively.

We assume that the phoretic mobility is uniform over the surface of the active colloid. 
Thus, the slip velocity follows readily from Eq.~\eqref{eq:vs} as
\begin{equation}
    \bm{v}_\mathrm{S} = 
    \frac{\mu q_1}{3R} 
    \left(  \hat{\bm{u}} + 
    \left( \bm{I} - \hat{\bm{n}}\hat{\bm{n}} \right) \cdot \hat{\bm{u}}
    \right) .
\end{equation}

By making use of Eqs.~\eqref{eq:transl_rot}, the resulting translational velocity in a bulk medium is obtained as
\begin{equation}
    \bm{V}_\mathrm{B} = -\frac{2\mu q_1}{3R} \, \hat{\bm{u}} \, .
    \label{eq:V_BULK}
\end{equation}
For a constant phoretic mobility in a bulk medium far from any boundaries, the active colloid exhibits purely translational motion along its axis of symmetry without reorientation.
We next examine the effect of the wedge walls.

\subsection{First reflection}

To determine the concentration field in the vicinity of the wedge, we employ the method of reflections. This method involves calculating the first reflection so as to satisfy the boundary conditions at the walls. Since this first reflection violates the boundary conditions at the surface of the active colloid, a second reflection is introduced to enforce the conditions there to leading order in the ratio of particle size to the distance from the nearest wall, following a far-field approach. This allows us to compute the wall-induced phoretic velocities. In the present work, we report only the contribution from diffusiophoresis and do not consider hydrodynamic effects.

In this manner, we express the solution for the concentration field in the form
\begin{equation}
    c = c^{(0)} + c^{(1)} + c^{(2)} \, .
\end{equation}
Furthermore, we decompose each of these three terms into source monopole and source dipole contributions, i.e.,
\begin{equation}
    c^{(n)} = c_\mathrm{M}^{(n)} + c_\mathrm{D}^{(n)} \, , \quad n \in \{0,1,2\} \, .
\end{equation}

Imposing no-flux boundary conditions at the walls implies that 
\begin{subequations}
    \begin{align}
    \left. \frac{\partial}{\partial\theta} \left( c_\mathrm{M}^{(0)} + c_\mathrm{M}^{(1)} \right) \right|_{\theta = \pm\alpha} = 0 \, , \label{eq:BCs_FIRST_REF_MONO} \\[3pt]
    \left. \frac{\partial}{\partial\theta} \left( c_\mathrm{D}^{(0)} + c_\mathrm{D}^{(1)} \right) \right|_{\theta = \pm\alpha} = 0 \, . \label{eq:BCs_FIRST_REF_DIPOLE}
\end{align}
\end{subequations}

\subsubsection{Source monopole}

We first consider the first reflection of the source monopole. As given above in Eq.~\eqref{eq:c0}, the source monopole contribution to the concentration field in an unbounded fluid medium under a constant flux boundary condition at the surface of the active particle is given by
\begin{equation}
    c_\mathrm{M}^{(0)} = {q_0}\, \frac{R}{s} \, .
\end{equation}

In cylindrical coordinates, the distance from the singularity is expressed as
\begin{equation}
    s = \sqrt{r^2+\rho^2-2\rho r \cos(\theta-\beta)+z^2} \, . \label{eq:s}
\end{equation}

The core idea of our approach is to first express the known bulk solution in FKL space and then determine the solution for the first reflection that satisfies the no-flux boundary conditions at the walls. The solution in real space is subsequently obtained via the inverse FKL transform.

The Fourier transform of $1/s$ is readily obtained as
\begin{equation}
    \mathscr{F} \left\{  \frac{1}{s} \right\} = 
    2 K_0 \left( |k|s_0\right) , 
\end{equation}
where 
\begin{equation}
    s_0 = s(z=0) = \sqrt{r^2+\rho^2-2\rho r \cos(\theta-\beta)} \, .
\end{equation}

We then use the classical Table of Integral Transforms by Erd{\'e}lyi \textit{et al.} \cite[p.~175]{erdelyi54} to obtain the FKL transform of $1/s$ as
\begin{equation}
    \mathscr{T}_{i\nu} \left\{ \frac{1}{s} \right\}
    = \Gamma \ch(a\nu) \, K_{i\nu} (|k|\rho) \, ,
    \label{eq:FKL-inv-s}
\end{equation}
where
\begin{equation}
    a = \pi - |\theta - \beta| \, , \quad
    \Gamma = 2\pi \nu^{-1} \csh(\pi\nu)\, .
    \label{eq:a_def}
\end{equation}

Thus, the FKL-transform of the free-space point source is obtained as
\begin{equation}
    \widetilde{c}_\mathrm{M}^{\,\,(0)}
    = \Gamma q_0 R \ch(a\nu) \, K_{i\nu} (|k|\rho) \, .
\end{equation}

Using the solution form given by Eq.~\eqref{eq:solution_form}, the solution after the first reflection can be written as 
\begin{equation}
    \widetilde{c}_\mathrm{M}^{\,\,(1)} = R \left( A_\mathrm{MS}^{(1)} \sh(\theta\nu) + A_\mathrm{MC}^{(1)} \ch(\theta\nu) \right) .
\end{equation}

By imposing zero-flux boundary conditions at the wedge surfaces $\theta=\pm\alpha$, as specified in \eqref{eq:BCs_FIRST_REF_MONO}, the coefficients $A_\mathrm{MS}^{(1)}$ and $A_\mathrm{MC}^{(1)}$ are determined as
\begin{equation}
    A_\mathrm{MX}^{(1)} = \Gamma q_0 \,\Lambda_\mathrm{MX}^{(1)} K_{i\nu}(|k| \rho) \, ,
\end{equation}
where $\mathrm{X} \in \{ \mathrm{S}, \mathrm{C} \}$.
Here, we have defined
\begin{subequations}
    \label{eq:Lam_M}
    \begin{align}
    \Lambda_\mathrm{MS}^{(1)} &= \sh(\beta \nu) \ch((\pi - \alpha)\nu) \sch(\alpha \nu)\, , \\
    \Lambda_\mathrm{MC}^{(1)} &= \ch(\beta \nu) \sh((\pi - \alpha)\nu) \csh(\alpha \nu) \, .
\end{align}
\end{subequations}

The real-space solution for the first reflection of the concentration field can be determined by applying the inverse FKL transform.
The integration over the axial wavenumber $k$ can be performed using tabulated integrals, resulting in a single remaining integral over the radial wavenumber $\nu$, which is then computed numerically using standard routines.
The solution can be written as
\begin{equation}
    c_\mathrm{M}^{(1)}(r,\theta,z) = q_0 \int_0^\infty \psi_\mathrm{M}^{(1)} (\theta,\nu) \, \mathcal{K}_{i\nu}(r,z) \, \mathrm{d}\nu \, , 
    \label{eq:cM1_sol}
\end{equation}
wherein the kernel function is defined by
\begin{equation}
    \mathcal{K}_{i\nu}(r,z)
    = \frac{4R}{\pi^2} \int_0^\infty 
    \cos(kz) 
    K_{i\nu}(kr) K_{i\nu}(k\rho) \, \mathrm{d}k \, ,
    \label{eq:Kernel}
\end{equation}
and 
\begin{equation}
    \psi_\mathrm{M}^{(1)} (\theta,\nu) = \Lambda_\mathrm{MS}^{(1)} \sh(\theta\nu) + \Lambda_\mathrm{MC}^{(1)} \ch(\theta\nu)\, .
    \label{eq:psiM}
\end{equation}

The improper integral in Eq.~\eqref{eq:Kernel} is convergent, and its value is reported in classical textbooks as
\begin{equation}
    \mathcal{K}_{i\nu} (r,z) = 
    \frac{R}{\sqrt{\rho r}} \,
    P_{i\nu-\frac{1}{2}} (w) \, \sch(\pi\nu) \, ,
    \label{eq:Kinu_result}
\end{equation}
see, for instance, Prudnikov \textit{et al.}~\cite[p.~390]{prudnikov1992integrals}, or Gradshteyn and Ryzhik~\cite[p.~719]{gradshteyn2014table}. Here, $P_n$ denotes the Legendre function of the first kind of degree $n$, with argument
\begin{equation}
    w = \frac{1}{2\rho r} \left( \rho^2+r^2+z^2 \right) .
    \label{eq:w}
\end{equation}

\begin{figure}
    \centering
    \includegraphics[width=0.85\linewidth]{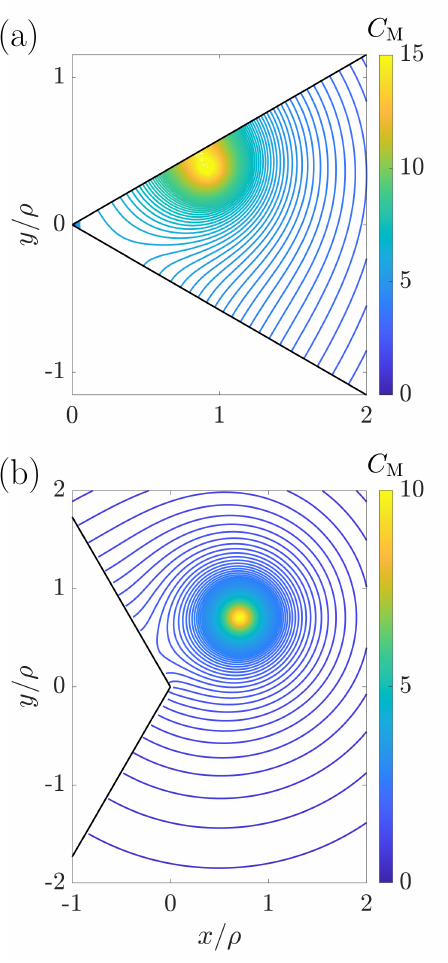}
    \caption{
    Contour plots of the scaled monopole concentration field around an active particle in the radial-azimuthal plane for
    (a) an obtuse wedge with $\alpha = \pi/6$ and the particle located at $\beta = \pi/8$, and
    (b) a salient wedge with $\alpha = 2\pi/3$ and the particle located at $\beta = \pi/4$.
    In both cases, the results are shown in the axial plane $z/\rho = 0.1$.
    }
    \label{fig:M_RThe}
\end{figure}

The first reflection is obtained by numerically integrating the infinite integral in Eq.~\eqref{eq:cM1_sol}. Together with the bulk solution, the zero-flux boundary conditions at the walls are satisfied. We define the scaled monopole concentration as
\begin{equation}
    C_\mathrm{M} = \frac{\rho}{q_0 R} \left( c_\mathrm{M}^{(0)} + c_\mathrm{M}^{(1)} \right) .
    \label{eq:C_M_scaled}
\end{equation}

In Fig.~\ref{fig:M_RThe}, we show example contour plots of isoconcentration lines around an active particle modeled as a source monopole. Results are presented for (a) an obtuse wedge with $\alpha = \pi/6$ and (b) a salient wedge with $\alpha = 2\pi/3$. The particle is positioned at an orientation $\beta = \pi/8$ in the obtuse wedge and $\beta = \pi/4$ in the salient wedge. 
Computations are performed in the radial-azimuthal plane at $z/\rho = 0.1$.
We observe that the presence of the wedge causes a significant distortion of the isoconcentration lines due to the no-flux boundary condition imposed at the walls. As a result, the isoconcentration lines become perpendicular to the wedge surfaces.
Far from the particle and the boundaries, the concentration gradually decays toward the background level, and the effect of the wedge geometry becomes negligible.

\begin{figure}
    \centering
    \includegraphics[width=0.9\linewidth]{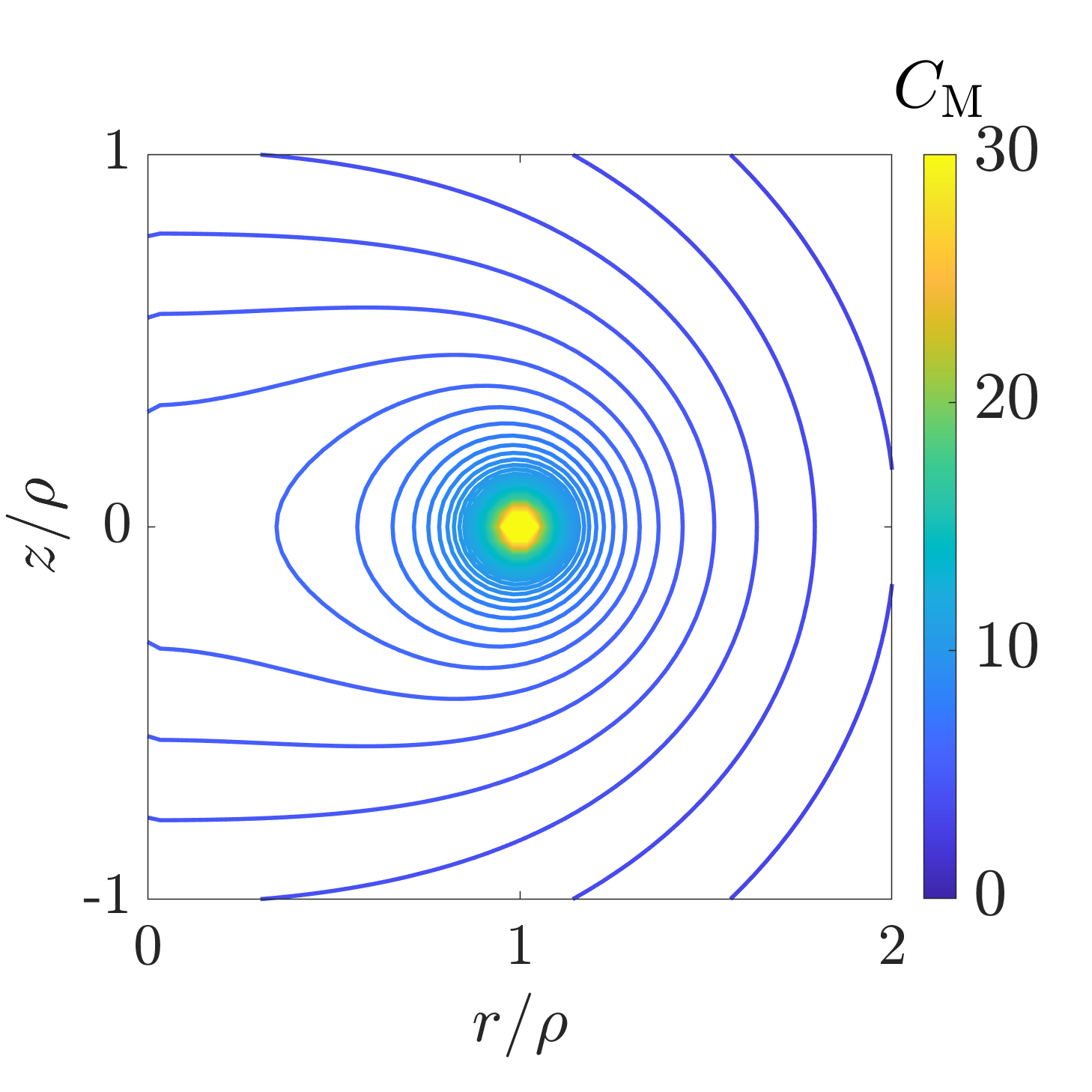}
    \caption{Contour plots of the scaled monopole concentration field around an active particle in the radial-axial plane for $\alpha = \pi/6$ and $\beta = \pi/12$. Results are shown in the azimuthal plane $\theta = 0$.}
    \label{fig:M_RZ}
\end{figure}

In Fig.~\ref{fig:M_RZ}, we present a contour plot of the monopole concentration in the radial-axial plane at $\theta = 0$. Results are shown for an obtuse wedge with $\alpha = \pi/6$ and a particle positioned at $\beta = \pi/12$. The isoconcentration lines in the right portion of the plot are only weakly affected by the wedge, reflecting the behavior of a source monopole in bulk. In contrast, near the wedge edge located at $r = 0$, the lines of equal concentration are strongly distorted due to the presence of the boundary.
Overall, the method described here demonstrates its robustness in determining the concentration field in the wedge geometry.

\subsubsection{Source dipole}

For the source dipole singularity, we proceed analogously by first expressing the bulk solution in FKL space and then matching the coefficients to satisfy the boundary conditions at the walls.

We express the unit orientation vector in terms of the variables $\delta$ and~$\lambda$, representing the polar and azimuthal angles in spherical coordinates, respectively.
Specifically,
\begin{equation}
    \hat{\bm{u}} = 
    \begin{pmatrix}
        \sin\delta \cos\lambda \\
        \sin\delta \sin\lambda \\
        \cos\delta 
    \end{pmatrix} ,
\end{equation}
where $\delta$ and $\lambda$ denote the polar and azimuthal angles, respectively.

The source dipolar contribution to the bulk concentration field, without the wedge, is given by [c.f.~Eq.~\eqref{eq:c0}]
\begin{equation}
    c_\mathrm{D}^{(0)} = q_1 \left( \frac{R}{s} \right)^2 \hat{\bm{u}} \cdot \hat{\bm{s}} \, . 
    \label{eq:c_D_bulk}
\end{equation}
The dot product can be written in a compact form as
\begin{equation}
    \hat{\bm{u}} \cdot \hat{\bm{s}} = 
    \frac{1}{s}
    \left( r\cos(\theta-\lambda)-\rho\cos(\beta-\lambda) \right)\sin\delta + \frac{z}{s} \cos\delta \, .
    \notag 
\end{equation}

To determine the FKL-transform, we need to express the free-space contribution given by Eq.~\eqref{eq:c_D_bulk} in terms of derivatives of $1/s$. For this purpose, we use the following two identities:
\begin{equation}
\frac{z}{s^3} = 
-\frac{\partial}{\partial z} \frac{1}{s} \, , \quad   
     \frac{\rho r}{s^3} \sin(a)
    =  \frac{\partial}{\partial a} \frac{1}{s} \, , 
    \label{eq:Elementary-identities}
\end{equation}
where $a$ is defined earlier in Eq.~\eqref{eq:a_def}.
Hence, 
\begin{equation}
    \hspace{-0.1cm}
    \frac{\hat{\bm{u}} \cdot \hat{\bm{s}}}{s^2} =
    \frac{\sin\delta}{\sin a}
    \left( \frac{\eta_1}{\rho} \, \cos(\theta-\lambda)
    +\eta_2 \cos(\beta-\lambda) \right)
    +\eta_3 \cos\delta \, , \notag 
\end{equation}
where 
\begin{equation}
    \eta_1 = \frac{\partial}{\partial a} \frac{1}{s} \, , \quad
    \eta_2 = -\frac{1}{r} \frac{\partial}{\partial a} \frac{1}{s} \, , \quad
    \eta_3 = -\frac{\partial}{\partial z} \frac{1}{s} \, .
\end{equation}

Thus, it remains only to determine the FKL transforms of $\eta_i$, with $i\in\{1,2,3\}$.

The FKL-transform of $\eta_1$ can be readily obtained by differentiating Eq.~\eqref{eq:FKL-inv-s} with respect to $a$ to give
\begin{equation}
    \widetilde{\eta}_1 = \Gamma \, \nu \sh(a\nu) \, K_{i\nu} (|k|\rho) \, .
\end{equation}

To determine the FKL-transform of $\eta_2$, we use the property of the FKL transform for a function divided by~$r$, as derived in Ref.~\cite{daddi2025proc}.
Specifically, 
\begin{equation}
 \widetilde{\eta}_2=
    -\mathscr{T}_{i\nu} \left\{ \frac{\eta_1}{r} \right\}
    = 
    \frac{|k|}{2i\nu} 
    \left( \mathscr{T}_{i\nu-1} \{\eta_1\} - \mathscr{T}_{i\nu+1} \{\eta_1\} \right) ,
    \label{eq:FKL-division-by-r}
\end{equation}
which leads to
\begin{equation}
\hspace{-0.4em}
    \widetilde{\eta}_2
    = \Gamma\left( \frac{\nu}{\rho}\cos a\sh(a\nu) + \sin a \ch(a\nu) \,\frac{\partial}{\partial \rho} \right)
    K_{i\nu}(|k|\rho) .
\end{equation}
Here, we have used the property of modified Bessel functions~\cite{abramowitz72}
\begin{equation}
    K_{i\nu\pm 1}(|k|\rho) = \frac{1}{|k|}
    \left( \pm \frac{i\nu}{\rho}\, K_{i\nu}(|k|\rho) - \frac{\partial}{\partial \rho} \right) K_{i\nu}(|k|\rho) \, .
\end{equation}

Finally, the FKL-transform of $\eta_3$ can be obtained using the derivative with respect to~$z$ as
\begin{equation}
    \widetilde{\eta}_3 = ik \mathscr{T}_{i\nu}
    \left\{ \frac{1}{s} \right\} 
    = ik \Gamma \ch(a\nu) \, K_{i\nu} (|k|\rho) \, .
\end{equation}

As a solution of Laplace’s equation, it can be expressed in the form given by Eq.~\eqref{eq:solution_form} as
\begin{equation}
    \widetilde{c}_\mathrm{D}^{\,\,(1)} = 
    R^2 \left(
    A_\mathrm{DS}^{(1)} \sh(\theta\nu) + A_\mathrm{DC}^{(1)} \ch(\theta\nu) \right) .
\end{equation}

By enforcing zero-flux at the wedge surfaces $\theta = \pm\alpha$, as specified in Eq.~\eqref{eq:BCs_FIRST_REF_DIPOLE}, the coefficients $A_\mathrm{DS}^{(1)}$ and $A_\mathrm{DC}^{(1)}$ are determined as
\begin{equation}
    A_\mathrm{DX}^{(1)} = \Gamma q_1
    \left( \frac{\Lambda_\mathrm{DX}^{(1)}}{\rho} + \mathrm{H}_\mathrm{DX}^{(1)} \frac{\partial}{\partial\rho} 
    + ik \, \Delta_\mathrm{DX}^{(1)} \right)
    K_{i\nu} (|k|\rho) \, , 
\end{equation}
where $\mathrm{X} \in \{ \mathrm{S}, \mathrm{C} \}$.
This brings the total number of coefficients to be determined to six.

\begin{figure}
    \centering
    \includegraphics[width=0.9\linewidth]{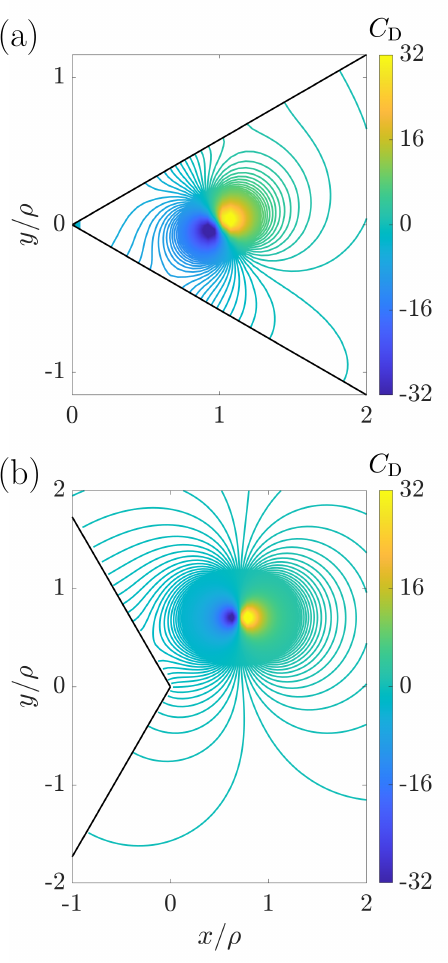}
    \caption{Contour plots of the scaled dipole concentration field around an active particle in the radial-azimuthal plane for (a)~an obtuse wedge with $\alpha = \pi/6$, where the particle is located at $\beta = 0$ and has angular orientation $(\lambda,\delta) = (\pi/6, \pi/2)$, and (b)~a salient wedge with $\alpha = 2\pi/3$, where the particle is located at $\beta = \pi/4$ and has angular orientation $(\lambda,\delta) = (0, \pi/2)$.
    In both cases, the results are shown in the axial plane $z/\rho = 0.1$.}
    \label{fig:D_RThe}
\end{figure}

In real space, the solution for the first reflection of the source dipole is given by
\begin{equation}
    c_\mathrm{D}^{(1)} = q_1 \int_0^\infty 
   \psi_\mathrm{D}^{(1)} (\theta, \nu) \,
    \mathcal{K}_{i\nu}(r,z) \, \mathrm{d}\nu \, ,
    \label{eq:cD1_sol}
\end{equation}
where
\begin{equation}
    \psi_\mathrm{D}^{(1)} = 
    R \left( 
    \frac{1}{\rho}\, \zeta_\mathrm{D}^{(1)}
    + \xi_\mathrm{D}^{(1)}  \, \frac{\partial}{\partial\rho}
    -\phi_\mathrm{D}^{(1)} \, \frac{\partial}{\partial z} \right) .
    \label{eq:psiD}
\end{equation}
Here, we have defined
\begin{subequations} \label{eq:zeta_xi_phi}
    \begin{align}
    \zeta_\mathrm{D}^{(1)} (\theta, \nu) &= \Lambda_\mathrm{DS}^{(1)} \sh(\theta\nu) + \Lambda_\mathrm{DC}^{(1)} \ch(\theta\nu) \, , \\[3pt]
    \xi_\mathrm{D}^{(1)} (\theta, \nu) &=
    \mathrm{H}_\mathrm{DS}^{(1)} \sh(\theta\nu) + \mathrm{H}_\mathrm{DC}^{(1)} \ch(\theta\nu) \, , \\[3pt]
    \phi_\mathrm{D}^{(1)} (\theta, \nu) &= \Delta_\mathrm{DS}^{(1)} \sh(\theta\nu) + \Delta_\mathrm{DC}^{(1)} \ch(\theta\nu) \, .
\end{align}
\end{subequations}
Note that the expression for the kernel function $\mathcal{K}_{i\nu}$ is provided in Eq.~\eqref{eq:Kinu_result}.
Defining
\begin{subequations}
    \begin{align}
    \sigma_1 &= \th(\alpha \nu) \sh(\pi \nu)-\ch(\pi \nu) \, , \\
    \sigma_2 &= \cth(\alpha \nu) \sh(\pi \nu) - \ch(\pi \nu) \, , 
\end{align}
\end{subequations}
the coefficients are obtained as
\begin{subequations} 
       \label{eq:coeffs_Dipole}
    \begin{align}
\Lambda_\mathrm{DS}^{(1)} &=\phantom{+} 
\nu \sigma_1 \sin\delta \ch(\beta \nu) \sin(\beta - \lambda) \, , \\
\Lambda_\mathrm{DC}^{(1)}&= -\nu\sigma_2
 \sin\delta \sh(\beta \nu) \sin(\beta - \lambda) \, , \\
\mathrm{H}_\mathrm{DS}^{(1)} &=-\sigma_1  
\sin\delta \sh(\beta \nu) \cos(\beta - \lambda)  \, , \\
\mathrm{H}_\mathrm{DC}^{(1)} &= \phantom{+} \sigma_2
\sin\delta \ch(\beta \nu) \cos(\beta - \lambda) \, , \\
\Delta_\mathrm{DS}^{(1)} &= -\sigma_1
\cos\delta \sh(\beta \nu) \, , \\
\Delta_\mathrm{DC}^{(1)} &= \phantom{+} \sigma_2
\cos\delta \ch(\beta \nu) \, .
\end{align}
\end{subequations}

We define the scaled dipole concentration as
\begin{equation}
    C_\mathrm{D} = \frac{1}{q_1} \left(\frac{\rho}{R} \right)^2 \left( c_\mathrm{D}^{(0)} + c_\mathrm{D}^{(1)} \right) .
    \label{eq:C_D_scaled}
\end{equation}

Figure~\ref{fig:D_RThe} shows contour plots of the scaled dipole concentration field around an active particle in the radial–azimuthal plane for (a) an obtuse wedge with $\alpha = \pi/6$, where the particle is positioned at $\beta = 0$ with angular orientation $(\lambda,\delta) = (\pi/6,\pi/2)$, and (b) a salient wedge with $\alpha = 2\pi/3$, where the particle is located at $\beta = \pi/4$ with angular orientation $(\lambda,\delta) = (0,\pi/2)$. In both cases, the fields are evaluated in the axial plane $z/\rho = 0.1$.
The source dipole arises from the superposition of two source monopoles—one acting as a source and the other as a sink—placed close to each other. These monopoles are aligned along $\pi/6$ in (a) and along the $\hat{\bm{x}}$ axis in (b).
Overall, the isoconcentration lines deviate markedly from the bulk behavior, reflecting the influence of the boundaries.
The streamlines are normal to the walls, consistent with the no-flux boundary condition.
Note that a source dipole aligned along the axial direction for $\delta = 0$ produces lines of equal concentration in the radial–azimuthal plane that closely resemble those of a source monopole and are therefore not shown here.

\begin{figure}
    \centering
    \includegraphics[width=0.9\linewidth]{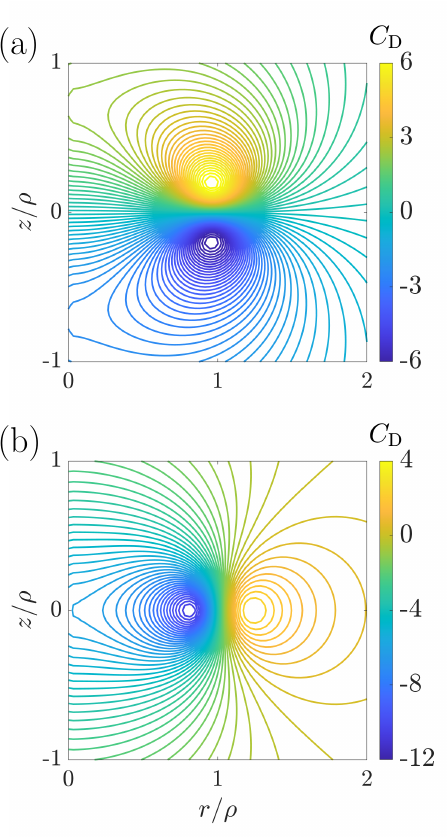}
    \caption{
    Contour plots of the scaled dipole concentration field around an active particle in the radial-axial plane for $\alpha = \pi/6$ and $\beta = \pi/12$, shown for the orientations (a) $\delta = 0$ and (b) $(\lambda,\delta) = (\pi/12, \pi/2)$. Results are displayed in the azimuthal plane $\theta = 0$.
   }
    \label{fig:D_RZ}
\end{figure}

Figure~\ref{fig:D_RZ} presents contour plots of the scaled dipole concentration field around an active particle in the radial–axial plane for $\alpha = \pi/6$ and $\beta = \pi/12$, shown for the orientations (a) $\delta = 0$ (parallel to the wedge axis) and (b) $(\lambda,\delta) = (\pi/12, \pi/2)$ (perpendicular to the wedge axis). The results are displayed in the azimuthal plane $\theta = 0$.
As before, the isoconcentration lines are more strongly affected in the region near the wedge.

\subsection{Second reflection}

To maintain the flux boundary condition at the surface of the active colloid to $\mathcal{O}(\epsilon^3)$ for the source monopole and to $\mathcal{O}(\epsilon^4)$ for the source dipole, an additional concentration field must be included. This is done by adding a field that is singular at the particle center such that the normal gradient of the first and second reflections of the concentration vanishes at the particle surface up to a given~$\epsilon$ order. Below, we determine this correction for both the monopole and dipole singularities.

We adopt a spherical coordinate system centered at the particle’s center defined as
\begin{equation}
    \hat{\bm{s}} =
    \begin{pmatrix}
        \sin \vartheta \cos\varphi \\
        \sin\vartheta \sin\varphi \\
        \cos\vartheta 
    \end{pmatrix} ,
\end{equation}
where $\vartheta$ and $\varphi$ denote the polar and azimuthal angles, respectively, in the particle-fixed frame.

Accordingly, the passage from cylindrical to spherical coordinates is made using the relations
\begin{subequations}
    \begin{align}
    r\cos\theta-\rho\cos\beta &= s\sin\vartheta\cos\varphi \, , \\
     r\sin\theta-\rho\sin\beta &= s\sin\vartheta\sin\varphi \, , \\
     z&= s\cos\vartheta \, .
\end{align}
\end{subequations}

This leads to the determination of the polar coordinates as
\begin{subequations} \label{eq:cylind_as_fct_sph}
    \begin{align}
    r &= \sqrt{\rho^2+s^2\sin^2\vartheta+2\rho s \cos(\beta-\varphi)\sin\vartheta} \, , \\
    \theta &= \arctan \left( \frac{s\sin\vartheta\sin\varphi + \rho\sin\beta}{s\sin\vartheta\cos\varphi + \rho\cos\beta} \right).
\end{align}
\end{subequations}

\subsubsection{Source monopole}

To satisfy the flux boundary condition on the surface of the active colloid, we require the normal gradients of the first and second reflections to vanish.
Specifically, 
\begin{equation}
    \left. \frac{\partial}{\partial s} 
    \left( c_\mathrm{M}^{(1)} + c_\mathrm{M}^{(2)} \right) \right|_{s=R} = \mathcal{O}\left( \epsilon^3 \right) \, .
    \label{eq:reflected_M}
\end{equation}

Using the solution for $c_\mathrm{M}^{(1)}$ from Eq.~\eqref{eq:cM1_sol}, together with the coordinate transformation equations in Eqs.~\eqref{eq:cylind_as_fct_sph}, we obtain, to leading order,
\begin{align}
    \left. \frac{\partial c_\mathrm{M}^{(1)}}{\partial s} \right|_{s=R} = 
    -\frac{q_0}{2} \, \epsilon^2 \sin^2(\alpha-\beta) 
    \int_0^\infty \bm{C}_\mathrm{M}^{(1)} \cdot \hat{\bm{s}} \, \mathrm{d}\nu
     \label{eq:matching_M1}
\end{align}
plus terms of order $\epsilon^3$, where
\begin{equation}
    \bm{C}_\mathrm{M}^{(1)} =
    \left( \bm{X}_\mathrm{MS}^{(1)}
    \sh(\beta \nu)
    +
   \bm{X}_\mathrm{MC}^{(1)}
    \ch(\beta \nu) \right) \sch(\pi\nu) \, , 
    \label{eq:CM1}
\end{equation}
where we have defined
\begin{subequations}
    \begin{align}
    \bm{X}_\mathrm{MS}^{(1)} &= 
    \begin{pmatrix}        2\nu\Lambda_\mathrm{MC}^{(1)}\sin\beta+\Lambda_\mathrm{MS}^{(1)} \cos\beta \\[3pt]
    \Lambda_\mathrm{MS}^{(1)} \sin\beta
    -2\nu\Lambda_\mathrm{MC}^{(1)} \cos\beta \\[3pt]
    0
    \end{pmatrix} , \\[5pt]
    \bm{X}_\mathrm{MC}^{(1)} &= 
    \begin{pmatrix}
        2\nu\Lambda_\mathrm{MS}^{(1)} \sin\beta+ \Lambda_\mathrm{MC}^{(1)} \cos\beta \\[3pt]
        \Lambda_\mathrm{MC}^{(1)} \sin\beta
    -2\nu\Lambda_\mathrm{MS}^{(1)} \cos\beta \\[3pt]
    0
    \end{pmatrix} .
\end{align}
\end{subequations}

To determine the second reflection needed to cancel this term to $\mathcal{O}(\epsilon^3)$, we express the concentration in the form
\begin{equation}
    c_\mathrm{M}^{(2)} = q_0 \epsilon^2
    R \, \bm{C}_\mathrm{M}^{(2)} \cdot \boldsymbol{\nabla} \, \frac{R}{s} \, ,
    \label{eq:c2M}
\end{equation}
where $\bm{C}_\mathrm{M}^{(2)}$ is a vector determined to ensure that Eq.~\eqref{eq:reflected_M} is satisfied.
Accordingly,
\begin{equation}
    \left. \frac{\partial c_\mathrm{M}^{(2)}}{\partial s} \right|_{s=R}
    = 2 \bm{C}_\mathrm{M}^{(2)} \cdot \hat{\bm{s}} \, .
    \label{eq:matching_M2}
\end{equation}

By substituting Eqs.~\eqref{eq:matching_M1} and \eqref{eq:matching_M2} into Eq.~\eqref{eq:reflected_M} and solving for $\bm{C}_\mathrm{M}^{(2)}$, we obtain
\begin{equation}
    \bm{C}_\mathrm{M}^{(2)} = 
    \frac{1}{4}\, \sin^2(\alpha-\beta) \int_0^\infty \bm{C}_\mathrm{M}^{(1)} \, \mathrm{d}\nu \, .
    \label{eq:CM2}
\end{equation}

\subsubsection{Source dipole}

We follow analogous mathematical steps to determine the second reflection of the source dipole concentration field. In this case, we require that
\begin{equation}
    \left. \frac{\partial}{\partial s} 
    \left( c_\mathrm{D}^{(1)} + c_\mathrm{D}^{(2)} \right) \right|_{s=R} = \mathcal{O}\left( \epsilon^4 \right) .
    \label{eq:reflected_D}
\end{equation}

The solution for the first reflection, $c_\mathrm{D}^{(1)}$, is given by Eq.~\eqref{eq:cD1_sol}, and its normal gradient, to leading order, is given by
\begin{align}
    \left. \frac{\partial c_\mathrm{D}^{(1)}}{\partial s} \right|_{s=R} = 
    -\frac{q_0}{8} \, \epsilon^3 \sin^3(\alpha-\beta) 
    \int_0^\infty \bm{C}_\mathrm{D}^{(1)} \cdot \hat{\bm{s}} \, \mathrm{d}\nu
     \label{eq:matching_D1}
\end{align}
plus terms of order $\epsilon^4$, where
\begin{equation}
    \bm{C}_\mathrm{D}^{(1)} = \left(
   \bm{X}_\mathrm{DS}^{(1)}
    \sh(\beta \nu)
    +
    \bm{X}_\mathrm{DC}^{(1)}
    \ch(\beta \nu) \right) \sch(\pi\nu) \, .
    \label{eq:CD1}
\end{equation}
Defining 
\begin{subequations}
    \begin{align}
    M_\mathrm{DX}^{(1)} &= 4\nu \left( H_\mathrm{DX}^{(1)}-2\Lambda_\mathrm{DX}^{(1)} \right) , \\
    N_\mathrm{DX}^{(1)} &= 4 \left( \nu^2 H_\mathrm{DX}^{(1)} -\Lambda_\mathrm{DX}^{(1)} \right) + 3H_\mathrm{DX}^{(1)} \, , 
\end{align}
\end{subequations}
where $X \in \{ \mathrm{S}, \mathrm{C} \}$, the column vectors $\bm{X}_\mathrm{DS}^{(1)}$ and $\bm{X}_\mathrm{DC}^{(1)}$ are given by 
\begin{subequations}
    \begin{align}
    \bm{X}_\mathrm{DS}^{(1)} &= 
    \begin{pmatrix}
        M_\mathrm{DC}^{(1)} \sin\beta + N_\mathrm{DS}^{(1)} \cos\beta \\[3pt]
         N_\mathrm{DS}^{(1)} \sin\beta
    - M_\mathrm{DC}^{(1)} \cos\beta \\[3pt]
    \left(4\nu^2+1 \right) \Delta_\mathrm{DS}^{(1)}
    \end{pmatrix} , \\[5pt]
    \bm{X}_\mathrm{DC}^{(1)} &= 
    \begin{pmatrix}
        M_\mathrm{DS}^{(1)} \sin\beta + N_\mathrm{DC}^{(1)} \cos\beta \\[3pt]
        N_\mathrm{DC}^{(1)} \sin\beta
    - M_\mathrm{DS}^{(1)} \cos\beta \\[3pt]
    \left(4\nu^2+1 \right) \Delta_\mathrm{DC}^{(1)}
    \end{pmatrix} .
\end{align}
\end{subequations}

To find the second reflection required to eliminate this term to $\mathcal{O}(\epsilon^4)$, we express the concentration in the form
\begin{equation}
    c_\mathrm{D}^{(2)} = q_1 \epsilon^3
    R \, \bm{C}_\mathrm{D}^{(2)} \cdot \boldsymbol{\nabla} \, \frac{R}{s} \, , 
    \label{eq:c2D}
\end{equation}
where $\bm{C}_\mathrm{D}^{(2)}$ is a vector to be determined subsequently.
Thus,
\begin{equation}
    \left. \frac{\partial c_\mathrm{D}^{(2)}}{\partial s} \right|_{s=R}
    = 2 \bm{C}_\mathrm{D}^{(2)} \cdot \hat{\bm{s}} \, .
    \label{eq:matching_D2}
\end{equation}

By substituting Eqs.~\eqref{eq:matching_D1} and \eqref{eq:matching_D2} into Eq.~\eqref{eq:reflected_D} and solving for $\bm{C}_\mathrm{D}^{(2)}$, we obtain
\begin{equation}
    \bm{C}_\mathrm{D}^{(2)} = 
    -\frac{1}{16}\, \sin^3(\alpha-\beta) \int_0^\infty \bm{C}_\mathrm{D}^{(1)} \, \mathrm{d}\nu \, .
    \label{eq:CD2}
\end{equation}

\section{Phoretic velocity}
\label{sec:phoretic_velocity}

\begin{figure}
    \centering
    \includegraphics[width=\linewidth]{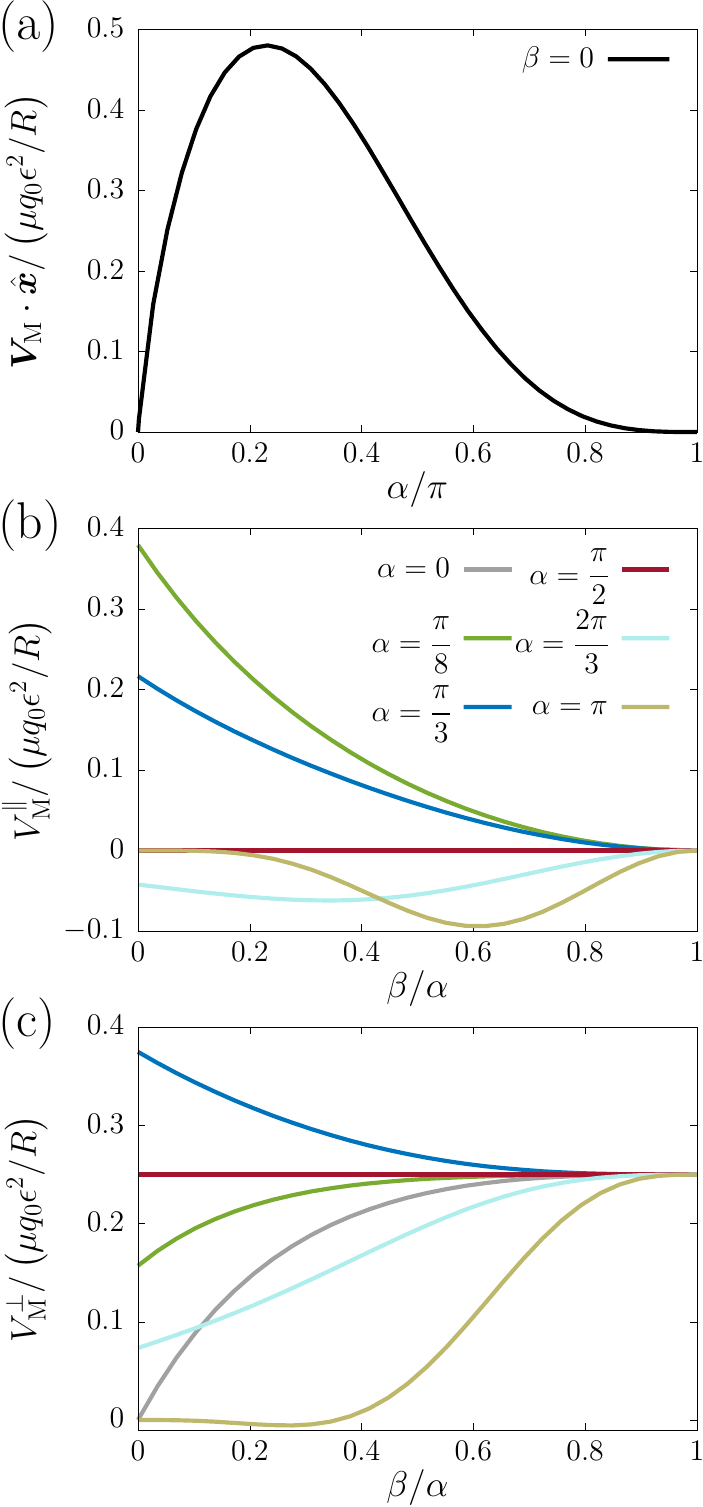}
    \caption{Scaled phoretic velocity induced by a source monopole near a wedge: (a) translational velocity along the~$\hat{\bm{x}}$ direction as a function of $\alpha$ in the case $\beta=0$, and variations of the (b) parallel and (c) normal components as functions of~$\beta/\alpha$ for different values of $\alpha$.}
    \label{fig:V_M}
\end{figure}

The self-induced phoretic velocity can now be determined using the solutions for the first and second reflections of the concentration field. The leading-order contributions from the monopole and dipole are expected to scale as~$\epsilon^2$ and~$\epsilon^3$, respectively. Notably, considering only the first reflection when determining the phoretic velocity introduces errors in the leading-order contributions (see Ref.~\cite{malgaretti2018self}).

The phoretic slip velocity is given by Eq.~\eqref{eq:vs}, with its expression in the particle-centered spherical coordinate system as
\begin{equation}
    \bm{v}_\mathrm{S} = 
    \frac{\mu}{R} \left. \left( \frac{\partial c}{\partial \vartheta}\, \hat{\bm{e}}_\vartheta 
    + \frac{1}{\sin\vartheta} \frac{\partial c}{\partial \varphi} \, \hat{\bm{e}}_\varphi
    \right) \right|_{s=R} ,
\end{equation}
which projects onto the Cartesian coordinate system as
\begin{equation}
    \bm{v}_\mathrm{S} = \frac{\mu}{R}
    \begin{pmatrix}
        \cfrac{\partial c}{\partial \vartheta}\, \cos\vartheta \cos\varphi - 
        \cfrac{\sin\varphi}{\sin\vartheta} \cfrac{\partial c}{\partial \varphi} \\[3pt]
        \cfrac{\partial c}{\partial \vartheta} \, \cos\vartheta \sin\varphi + \cfrac{\cos\varphi}{\sin\vartheta} \cfrac{\partial c}{\partial \varphi} \\[3pt]
        -\cfrac{\partial c}{\partial \vartheta} \, \sin\vartheta
    \end{pmatrix}_{s=R} .
    \label{vs_final}
\end{equation}

The translational and rotational phoretic velocities are obtained by substituting the concentration field into the slip velocity expression in Eq.~\eqref{vs_final} and applying Eqs.~\eqref{eq:transl_rot}.
The surface integration in spherical coordinates is given by
\begin{equation}
    \langle \cdot \rangle = 
    \frac{1}{4\pi} \int_0^{2\pi} \int_0^\pi (\cdot) \sin\vartheta \, \mathrm{d}\vartheta \,\mathrm{d}\varphi \, . 
\end{equation}

We denote by $\bm{V}_\mathrm{M}$ and $\bm{V}_\mathrm{D}$ the contributions arising from the monopole and dipole singularities, respectively.

\subsection{Source monopole}

The monopole contribution is obtained as 
\begin{align}
    \bm{V}_\mathrm{M} = 
    \frac{\mu q_0}{2R} \, \epsilon^2 \sin^2(\alpha-\beta)
    \int_0^\infty \bm{C}_\mathrm{M}^{(1)} \, \mathrm{d}\nu
     \label{eq:V_M}
\end{align}
plus terms of order $\epsilon^3$, where the expression of $\bm{C}_\mathrm{M}^{(1)}$ is given in Eq.~\eqref{eq:CM1}.

By substituting the expressions of $\Lambda_\mathrm{MS}^{(1)}$ and $\Lambda_\mathrm{MC}^{(1)}$ from Eqs.~\eqref{eq:Lam_M} into Eq.~\eqref{eq:CM1}, we obtain
\begin{equation}
    \bm{C}_\mathrm{M}^{(1)} =
M_1 \begin{pmatrix}
    \cos\beta \\
    \sin\beta \\
    0
\end{pmatrix}
+ 
M_2 \begin{pmatrix}
    \sin\beta \\
    -\cos\beta \\
    0
\end{pmatrix} , \label{eq:CM1_vec}
\end{equation}
where 
\begin{subequations}
    \begin{align}
    M_1 &= \th(\pi\nu) \csh(2\alpha\nu)  \left( \ch(2\alpha\nu) + \ch(2\beta\nu) \right) - 1 \, , \\
    M_2 &= 2\nu \th(\pi\nu) \csh(2\alpha\nu)
    \sh(2\beta\nu)  \, .
\end{align}
\end{subequations}

In particular, in the special case $\beta=0$, corresponding to a particle located equidistant from the two walls, symmetry dictates that the particle moves solely along the $\hat{\bm{x}}$ direction. In this case, a simpler analytical expression for the induced velocity can be obtained as
\begin{equation}
    \bm{V}_\mathrm{M} = 
    \frac{\mu q_0 \epsilon^2}{R} \, 
 f_\mathrm{M}(\alpha) \, \hat{\bm{x}} \, ,
\end{equation}
where $f_\mathrm{M}(\alpha)$ depends only on the semi-opening angle $\alpha$ and is given by an infinite integral of the form
\begin{equation}
    f_\mathrm{M}(\alpha) = 
    \frac{1}{2} \, 
    \sin^2\alpha \int_0^\infty 
    \left( \cth(\alpha\nu) \th(\pi\nu) - 1 \right)
    \mathrm{d}\nu \, .
    \label{eq:f_M_integral}
\end{equation}
As shown in the Appendix, this integral can be written as a double infinite series
\begin{equation}
    f_\mathrm{M}(\alpha) =
    \sin^2\alpha
    \sum_{m,n} {f_\mathrm{M}}_{mn} \, ,
\end{equation}
where the summations run over all natural numbers $n=0,1,\dots$ and $m=0,1,\dots$, and
\begin{equation}
    {f_\mathrm{M}}_{mn} = \frac{(-1)^n }{2}
    \left( \frac{1}{A_{mn}}-
    \frac{1}{B_{mn}}
    \right) ,
\end{equation}
with
\begin{equation}
    A_{mn} = \pi n+\alpha m+\alpha \, , \quad
    B_{mn} = \pi n+\alpha m+\pi \, .
    \label{eq:Amn_Bmn}
\end{equation}
This double series provides accurate values for arbitrary~$\alpha$.
For commensurate angles of the form $\alpha_q = \pi/q$ with integer~$q$, the series reduces to a simple finite sum
\begin{equation}
    f_\mathrm{M}(\alpha_q) = \frac{1}{4}\, \sin^2 \alpha_q \sum_{k=1}^{q-1} \csc (k\alpha_q) \, .
    \label{eq:f_M_commensurate}
\end{equation}

We summarize in Tab.~\ref{tab:f_M} the exact values of $f_\mathrm{M}(\alpha_q)$ for several common commensurate angles.

\begin{table}[]
\centering
\renewcommand{\arraystretch}{2.75}
\setlength{\tabcolsep}{8pt}
\begin{tabular}{|c|c|c|c|c|c|}
\hline
$q = \cfrac{\pi}{\alpha_q}$ & 1 & 2 & 3 & 4 & 6 \\
\hline
$f_\mathrm{M}(\alpha_q)$ & 0 & $\cfrac{1}{4}$ & $\cfrac{\sqrt{3}}{4}$ & $\cfrac{1}{8} + \cfrac{\sqrt{2}}{4}$ & $\cfrac{5}{16} + \cfrac{\sqrt{3}}{12}$ \\
\hline
\end{tabular}
\caption{Exact values of the function $f_\mathrm{M}(\alpha_q)$ defined by Eq.~\eqref{eq:f_M_commensurate} for several common commensurate angles $\alpha_q = \pi/q$.
}
\label{tab:f_M}
\end{table}

In Fig.~\ref{fig:V_M} we show the variation of the self-induced phoretic velocity generated by a source monopole, obtained analytically from the integral expression in Eq.~\eqref{eq:V_M}. 
It is worth noting that the limit $\alpha \to$~0 corresponds to diffusiophoretic motion between two infinitely extended walls, whereas the limit $\alpha \to \pi$ corresponds to motion near a two-dimensional plane defined by $x \le 0$ and $y = 0$, extending infinitely along the $z$ direction.
Figure~\ref{fig:V_M}~(a) presents results for $\beta=0$ as a function of $\alpha$, corresponding to one-dimensional particle motion along the $\hat{\bm{x}}$ direction, with motion away from the wedge for $\mu q_0>0$ and toward the wedge for $\mu q_0<0$. The induced velocity vanishes in the limiting cases $\alpha=0$ and $\alpha=\pi$, and attains a maximum at an intermediate angle $\alpha/\pi \approx 0.23$. Notably, for $\alpha=\pi/2$ we recover the leading-order behavior near a planar wall, expressed as a scaled velocity normal to the wall of 0.25~\cite{ibrahim15,ibrahim2016walls, yariv16}; refer to Tab.~\ref{tab:f_M} for the case $q=2$.

Owing to the geometrical symmetry of the wedge, we restrict attention to $\beta \in [0,\alpha)$, so that the particle is located closer to the wedge at $\theta=\alpha$. Since monopole interactions induce only in-plane motion, here confined to the plane $z=0$, we define the components of the induced phoretic velocity parallel and perpendicular to the upper wall as
\begin{subequations} \label{eq:V_M_para_Perp}
    \begin{align}
    V_\mathrm{M}^\parallel &= \bm{V}_\mathrm{M} \cdot \hat{\bm{x}} \, \cos\alpha 
    + \bm{V}_\mathrm{M} \cdot \hat{\bm{y}} \, \sin\alpha \, , \\
    V_\mathrm{M}^\perp &= \bm{V}_\mathrm{M} \cdot \hat{\bm{x}} \, \sin\alpha 
    - \bm{V}_\mathrm{M} \cdot \hat{\bm{y}} \, \cos\alpha \, .
\end{align}
\end{subequations}

In Fig.~\ref{fig:V_M}~(b) and (c) we show the parallel and perpendicular components of the phoretic velocity as functions of $\beta/\alpha$ for several values of $\alpha$. 
As can be seen in Fig.~\ref{fig:V_M}~(a), the parallel component vanishes for $\alpha=0$ and $\alpha=\pi/2$, as expected from symmetry in these axisymmetric configurations. It also vanishes in the limit $\beta \to \alpha$, which effectively corresponds to a particle interacting with a single wall, since the influence of the second wall becomes negligible. For $\mu q_0>0$, we find that $V_\mathrm{M}^\parallel > 0$ for an obtuse wedge $(0<\alpha<\pi/2)$ and $V_\mathrm{M}^\parallel < 0$ for a salient wedge $(\pi/2<\alpha<\pi)$, indicating that the particle tends to move away from the wedge edge. This behavior is reversed for $\mu q_0<0$.
Figure~\ref{fig:V_M}~(b) shows the corresponding results for the component normal to the upper wall. In this case, the curves converge to the planar-wall limit as $\beta \to \alpha$, which corresponds to the configuration $\alpha=\pi/2$, where the scaled normal velocity is 0.25, independent of~$\beta$.

\subsection{Source dipole}

\begin{figure}
    \centering
    \includegraphics[width=\linewidth]{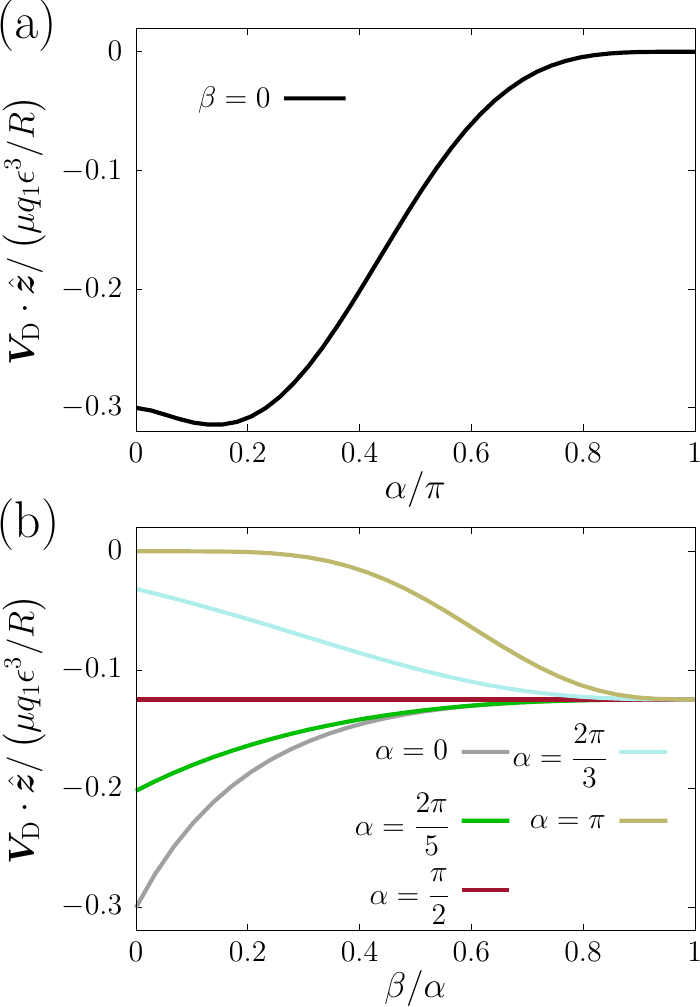}
    \caption{Axial component of the scaled phoretic velocity arising from source dipolar interactions for $\delta=0$: (a) variation with $\alpha$ for $\beta=0$, and (b) variation with $\beta/\alpha$ for different values of $\alpha$. }
    \label{fig:V_D1}
\end{figure}

The dipole contribution is obtained as 
\begin{align}
    \bm{V}_\mathrm{D} = 
    -\frac{\mu q_1}{8R} \, \epsilon^3 \sin^3(\alpha-\beta)
    \int_0^\infty \bm{C}_\mathrm{D}^{(1)} \, \mathrm{d}\nu
     \label{eq:V_D}
\end{align}
plus terms of order $\epsilon^4$.

By substituting the coefficient expressions from Eqs.~\eqref{eq:coeffs_Dipole} into Eq.~\eqref{eq:CD1}, we obtain
\begin{equation}
    \bm{C}_\mathrm{D}^{(1)} =
    \left( 
    D_0 \, \hat{\bm{e}}_0
    + D_1 \,
    \hat{\bm{e}}_1
    + 
    D_2 
    \hat{\bm{e}}_2 \,
    \right) \sin\delta 
    +
    D_3 \cos\delta \, \hat{\bm{z}} \, ,
    \label{eq:CD1_vec}
\end{equation}
where we have introduced, for convenience, the unit vectors $\hat{\bm{e}}_0 = \left(\cos\lambda, \sin\lambda, 0\right)^\top $, $\hat{\bm{e}}_1 = \left( \cos(2\beta-\lambda), \sin(2\beta-\lambda), 0 \right)^\top$, and $\hat{\bm{e}}_2 = \left( \sin(2\beta-\lambda), -\cos(2\beta-\lambda),  0 \right)^\top $. 

Defining the abbreviations 
\begin{subequations}
    \begin{align}
    Q_1 &= \th(\pi\nu)\csh(2\alpha\nu)\ch(2\nu\beta) \, \\
    Q_2 &= 1-\th(\pi\nu)\cth(2\alpha\nu) \, , 
\end{align}
\end{subequations}
the coefficients $D_0$ and $D_1$ are given by
\begin{subequations}
    \begin{align}
    D_0 &= \phantom{+}  Q_1 N_+ + Q_2 N_-  \, , \\[3pt]
    D_1 &= - Q_2 N_+ -Q_1 N_- 
    \, ,
\end{align}
\end{subequations}
where $N_+ = 3\left( 2\nu^2+1/2\right)$ and $N_- = 2\nu^2-3/2$.
In addition, we have defined $ D_2 = 2 M_2$ and $D_3 = \left(4\nu^2+1 \right) M_1.$

It is worth noting that the contribution to the phoretic velocity from the first reflection is twice the contribution from the second reflection. Accordingly, neglecting the second reflection results in relative errors of 1/3.
This led to an error in Ref.~\cite{malgaretti2018self} when determining the phoretic velocity near a planar fluid–fluid interface.

In the special case $\beta=0$, with the particle located midway between the two walls, and for $\delta=0$, the induced velocity from dipolar interactions has a single component along the~$\hat{\bm{z}}$~direction and can be expressed in a simplified analytical form
\begin{equation}
    \bm{V}_\mathrm{D} = 
    -\frac{\mu q_1 \epsilon^3}{R} \, 
 f_\mathrm{D}(\alpha) \, \hat{\bm{z}} \, ,
\end{equation}
where $f_\mathrm{D}(\alpha)$ depends solely on $\alpha$ and is given by an infinite integral of the form
\begin{equation}
    f_\mathrm{D}(\alpha) = 
    \frac{1}{8} \, 
    \sin^3\alpha \int_0^\infty
    \left( 4\nu^2+1 \right)
    \left( \cth(\alpha\nu) \th(\pi\nu) - 1 \right)
    \mathrm{d}\nu \, .
    \label{eq:f_D_integral}
\end{equation}
As shown in the Appendix, this integral can be evaluated and expressed as a double series of the form
\begin{equation}
    f_\mathrm{D}(\alpha) =
    \sin^3\alpha
    \sum_{m,n} {f_\mathrm{D}}_{mn} \, ,
\end{equation}
where 
\begin{equation}
    {f_\mathrm{D}}_{mn} = \frac{(-1)^n }{8}
    \left(  \frac{1}{A_{mn}} - \frac{1}{B_{mn}}
    + \frac{2}{A_{mn}^3}-\frac{2}{B_{mn}^3}
    \right) , 
\end{equation}
where the series coefficients $A_{mn}$ and $B_{mn}$ are defined earlier in Eqs. \eqref{eq:Amn_Bmn}.
For a commensurate semi-opening angle $\alpha_q=\pi/q$, analytical progress can be made, yielding a finite-series representation of the integral as
\begin{equation}
    f_\mathrm{D}(\alpha_q) = \frac{1}{8}\, \sin^3\alpha_q 
    \sum_{k=1}^{q-1} \csc^3(k\alpha_q) \, .
    \label{eq:f_D_commensurate}
\end{equation}

\begin{table}[]
\centering
\renewcommand{\arraystretch}{2.75}
\setlength{\tabcolsep}{8pt}
\begin{tabular}{|c|c|c|c|c|c|}
\hline
$q = \cfrac{\pi}{\alpha_q}$ & 1 & 2 & 3 & 4 & 6 \\
\hline
$f_\mathrm{D}(\alpha_q)$ & 0 & $\cfrac{1}{8}$ & $\cfrac{1}{4}$ & $\cfrac{1}{4}+\cfrac{\sqrt{2}}{32}$ & $\cfrac{17}{64}+\cfrac{\sqrt{3}}{36}$ \\
\hline
$g_\mathrm{D}(\alpha_q)$ & 0 & $\cfrac{1}{4}$ & $\cfrac{7}{16}$ & $\cfrac{3}{8} + \cfrac{\sqrt{2}}{16}$ & $\cfrac{11}{32} + \cfrac{7\sqrt{3}}{144}$ \\
\hline
$h_\mathrm{D}(\alpha_q)$ & 0 & $\cfrac{1}{8}$ & $\cfrac{5}{48} + \cfrac{9 \sqrt{3}}{32\, \pi}
$ & $\cfrac{3}{8} - \cfrac{\sqrt{2}}{32}$ & $\cfrac{29}{64} - \cfrac{5 \sqrt{3}}{144}$ \\
\hline
\end{tabular}
\caption{Exact values of the functions $f_\mathrm{D}(\alpha_q)$, $g_\mathrm{D}(\alpha_q)$, and $h_\mathrm{D}(\alpha_q)$, defined by Eqs.~\eqref{eq:f_D_commensurate}, \eqref{eq:g_D_commensurate}, and \eqref{eq:h_D_commensurate}, respectively, for several common commensurate angles $\alpha_q = \pi/q$.
}
\label{tab:fgh_D}
\end{table}

For $\beta=0$ and $\delta=\pi/2$, corresponding to motion in the $xy$ plane, the induced dipolar velocity can be written in the form
\begin{equation}
    \bm{V}_\mathrm{D} = -\frac{\mu q_1 \epsilon^3}{R}
    \left( g_\mathrm{D}(\alpha) \cos\lambda \, \hat{\bm{x}}
    +
    h_\mathrm{D}(\alpha) \sin\lambda \, \hat{\bm{y}}
    \right) ,
\end{equation}
where $g_\mathrm{D}(\alpha)$ and $h_\mathrm{D}(\alpha)$ depend only on $\alpha$ and are defined through infinite integrals
\begin{equation}
     g_\mathrm{D}(\alpha) = 
    \frac{1}{8} \, 
    \sin^3\alpha \int_0^\infty 
    \left(4\nu^2+3 \right)
    \left( \cth(\alpha\nu) \th(\pi\nu) - 1 \right)
    \mathrm{d}\nu \, ,
    \label{eq:g_D_integral}
\end{equation}
and
\begin{equation}
    h_\mathrm{D}(\alpha) = 
    \sin^3\alpha \int_0^\infty
    \nu^2 \left( 1-\th(\alpha\nu)\th(\pi\nu) \right) \mathrm{d}\nu \, .
    \label{eq:h_D_integral}
\end{equation}
Both integrals can be evaluated and written in terms of convergent double sums of the form; see Appendix for the derivation
\begin{equation}
    g_\mathrm{D}(\alpha) =
    \sin^3\alpha
    \sum_{m,n} {g_\mathrm{D}}_{mn} \, ,
\end{equation}
where 
\begin{equation}
    {g_\mathrm{D}}_{mn} = \frac{(-1)^n }{8}
    \left(  \frac{3}{A_{mn}} - \frac{3}{B_{mn}}
    + \frac{2}{A_{mn}^3}-\frac{2}{B_{mn}^3}
    \right) .
\end{equation}
For commensurate values of $\alpha_q=\pi/q$, $g_\mathrm{D}(\alpha_q)$ can be expressed as a finite series of the form
\begin{equation}
    g_\mathrm{D}(\alpha_q) = 
    \frac{1}{8}\, \sin^3\alpha_q 
    \sum_{k=1}^{q-1} \csc(k\alpha_q)
    \left( 1+\csc^2(k\alpha_q)\right)
    \, .
    \label{eq:g_D_commensurate}
\end{equation}

The integral $h_\mathrm{D}(\alpha)$ can be evaluated as
\begin{equation}
    h_\mathrm{D}(\alpha) = \sin^3\alpha 
    \left( \frac{3}{8}\, \zeta(3) \left( \frac{1}{\pi^3} + \frac{1}{\alpha^3} \right) 
    + \sum_{m,n}^\infty
    {h_\mathrm{D}}_{mn} 
    \right) ,
\end{equation}
where $\zeta$ denotes the Riemann zeta function, with $\zeta(3) \approx 1.2021$, and
\begin{equation}
    {h_\mathrm{D}}_{mn} = 
    \frac{(-1)^{n+m+1}}{\left( \pi n+\alpha m+\pi+\alpha \right)^3} \, .
\end{equation}
For commensurate angles $\alpha_q$ with even $q$, $h_\mathrm{D}(\alpha_q)$ can be written as a finite series of the form
\begin{equation}
\hspace{-0.2cm}
    h_\mathrm{D}(\alpha_q) = 
    \frac{1}{12}\, \sin^3\alpha_q \left(
    \frac{q^2}{2\pi} \left( 1-(-1)^q \right)
    +\sum_{k=1}^{\lfloor q/2\rfloor} H_{qk} \right) ,
    \label{eq:h_D_commensurate}
\end{equation}
where
\begin{equation}
    H_{qk} = \frac{(2k-1)(2k-q-1)(2k-2q-1)}{q}\tan\left(\frac{(2k-1)\alpha_q}{2}\right) . \notag 
\end{equation}
Here, $\lfloor \cdot \rfloor$ denotes the floor function, which ensures that the sum terminates at $k = q/2$ for even $q$ and at $k = (q-1)/2$ for odd $q$. In particular, Eq.~\eqref{eq:h_D_commensurate} simplifies when $q$ is even as
\begin{equation}
\hspace{-0.2cm}
    h_\mathrm{D}(\alpha_q) = 
    \frac{1}{8}\, \sin^3\alpha_q 
    \sum_{k=1}^{q-1} (-1)^k \csc(k\alpha_q)
    \left( 1-2\csc^2(k\alpha_q)\right) .
\end{equation}
For odd $k$, Eq.~\eqref{eq:h_D_commensurate} can be rewritten as
\begin{equation}
\hspace{-0.25cm}
h_\mathrm{D}(\alpha_q) = 
\frac{q}{6} \sin^3\alpha_q \left( \frac{q}{2\pi} + \sum_{k=1}^{ \frac{q-1}{2} } k \left( 1 - \frac{4k^2}{q^2} \right) \cot (k\alpha_q) \right).
\end{equation}
Tab.~\ref{tab:fgh_D} lists the values of $f_\mathrm{D}(\alpha_q)$, $g_\mathrm{D}(\alpha_q)$, and $h_\mathrm{D}(\alpha_q)$ for several common commensurate semi-opening angles $\alpha_k$.

In Fig.~\ref{fig:V_D1} we plot the variation of the scaled phoretic velocity along the axial direction of motion arising from source dipolar interactions with orientation vector $\hat{\bm{u}} \equiv \hat{\bm{z}}$, corresponding to $\delta=0$. Such a velocity represents a correction due to the presence of the wedge and must be added to the bulk contribution given by Eq.~\eqref{eq:V_BULK}. Figure~\ref{fig:V_D1}~(a) presents results for $\beta=0$ over the full range of $\alpha$. The velocity approaches the two-wall limit as $\alpha \to 0$, increases slightly in magnitude to reach an extremum at $\alpha/\pi \approx 0.14$, and then decreases to vanish in the limit $\alpha \to \pi$. The axial correction is positive for $\mu q_1<0$ and negative otherwise, and therefore has the same sign as the bulk contribution.
The limit $\alpha \to 0$ yields a scaled dipolar contribution of 0.3 in magnitude, corresponding to a phoretic particle confined to the midplane between two planar walls.

In Fig.~\ref{fig:V_D1}~(b) we show the axial component of the source dipolar contribution to the phoretic velocity as a function of~$\beta/\alpha$ for various values of $\alpha$, covering both obtuse and salient wedge configurations. The scaled axial velocity reaches $-0.125$, corresponding to the planar-wall limit at $\alpha=\pi/2$, independent of $\beta/\alpha$. The source dipole contribution to the axial velocity has the same sign as the bulk contribution for all values of $\alpha$. Notably, the velocity magnitude is larger for obtuse wedges than for salient ones, with all curves converging to the planar-wall limit as $\beta \to \alpha$.

\begin{figure}
    \centering
    \includegraphics[width=\linewidth]{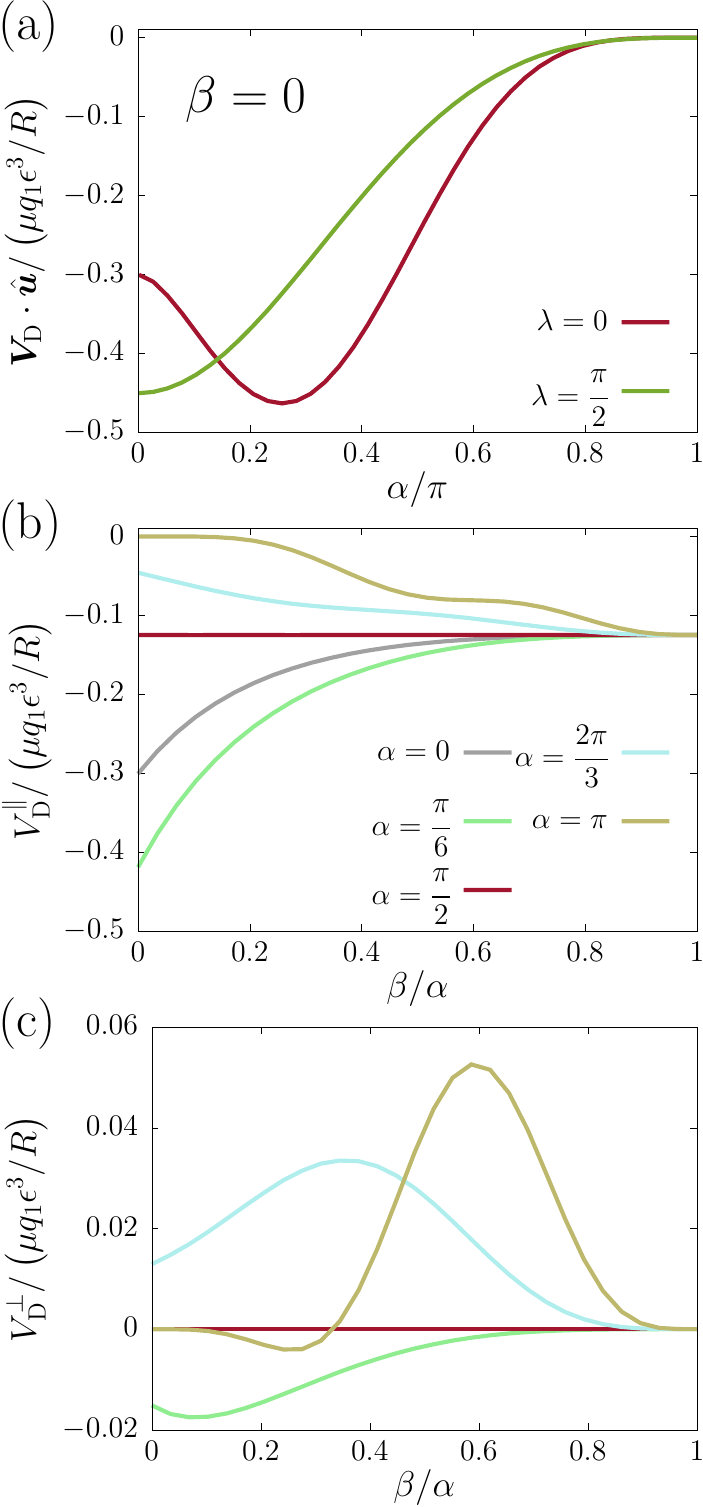}
    \caption{Variation of the scaled in-plane velocity arising from source dipolar interactions for $\delta=\pi/2$: (a) dependence on $\beta=0$ for $\lambda=0$ and $\lambda=\pi/2$, and (b) evolution with $\beta/\alpha$ for $\lambda=\alpha$ for different values of $\alpha$.  
   }
    \label{fig:V_D2}
\end{figure}

For an orientation vector lying in the $xy$ plane and aligned normal to the wedge edge, corresponding to $\delta=\pi/2$, the induced source dipolar contribution to the phoretic velocity depends on the azimuthal angle $\lambda$. 

The resulting in-plane phoretic velocity is shown in Fig.~\ref{fig:V_D2}~(a) for $\lambda = 0$ and $\lambda = \pi/2$, focusing on the case $\beta=0$. In both cases, the motion is one-dimensional: along the $\hat{\bm{x}}$ direction for $\lambda = 0$ and along the $\hat{\bm{y}}$ direction for $\lambda = \pi/2$. The dipolar contribution has the same sign as the bulk velocity in both cases. For $\lambda = 0$, the velocity exhibits a non-monotonic dependence on $\alpha$, initially increasing in magnitude to reach a maximum around $\alpha/\pi \approx 0.26$ before vanishing as $\alpha \to \pi$. In this configuration, the active particle moves toward the wedge vertex if $\mu q_1 < 0$ and away from it otherwise.
For $\lambda = \pi/2$, the dipolar contribution to the phoretic velocity decreases in magnitude. In this configuration, the active particle moves toward the lower wedge if $\mu q_1 < 0$ and toward the upper wedge otherwise. The scaled contributions in the limit $\alpha \to 0$ correspond to the solution for a particle confined to the midplane between two planar walls, taking values of $0.3$ and $0.45$ for $\lambda = 0$ and $\lambda = \pi/2$, respectively.

To examine the effect of particle position on the induced phoretic velocity, we focus here on the case $\lambda = \alpha$, where the particle is aligned parallel to the upper wedge. Figures~\ref{fig:V_D2}~(b) and~(c) show the corresponding parallel and perpendicular components of the velocity as functions of~$\beta/\alpha$ for various values of~$\alpha$. These components are defined as for the monopole contribution in Eqs.~\eqref{eq:V_M_para_Perp}.
For the parallel component in Fig.~\ref{fig:V_D2}~(b), all curves converge to $-0.125$ in the limit $\beta \to \alpha$, corresponding to $\alpha = \pi/2$. The velocity magnitude is generally larger for obtuse wedges than for salient wedges. The perpendicular component in Fig.~\ref{fig:V_D2}~(c) exhibits nontrivial behavior, including a possible sign change at intermediate values of $\beta$. This component vanishes for both $\alpha = 0$ and $\alpha = \pi/2$, as expected from symmetry.
All in all, we have successfully recovered the appropriate limiting behaviors and provided insight into the transition between them.

\section{Solution in the planar-wall limit}
\label{sec:planar_boundary}

We now demonstrate how the solution derived here coincides with previous results for a planar wall by setting $\alpha = \pi/2$. In this special case, the infinite integrals over the radial wavenumber can be expressed in closed analytical form, unlike the general case for arbitrary $\alpha$. We compare both the first and second reflections of the concentration, as well as the phoretic velocity induced by monopole and dipole interactions.

\subsection{Source monopole}

By setting $\alpha = \pi/2$ in Eqs.~\eqref{eq:Lam_M}, the coefficients corresponding to the source monopole reduce to $\Lambda_\mathrm{MS}^{(1)} = \sh(\beta\nu)$ and $\Lambda_\mathrm{MC}^{(1)} = \ch(\beta\nu)$, so that from Eq.~\eqref{eq:psiM} we have $\psi_\mathrm{M}^{(1)}(\theta, \nu) = \ch\big((\theta+\beta)\nu\big)$. The first reflection of the concentration field then follows directly from Eq.~\eqref{eq:cM1_sol} as
\begin{equation}
    c_\mathrm{M}^{(1)} = \frac{q_0 R}{\sqrt{\rho r}}
    \int_0^\infty 
    \ch \left( (\theta+\beta)\nu \right) \sch(\pi\nu) P_{i\nu-\frac{1}{2}}(w) \, \mathrm{d}\nu \, .
\end{equation}
We recall that $w$ is defined in Eq.~\eqref{eq:w}.

Defining the improper integral
\begin{equation}
    \mathcal{L} =
    \frac{1}{\sqrt{\rho r}}
    \int_0^\infty \ch ( b\nu) \sch(\pi\nu) \, P_{i\nu-\frac{1}{2}}(w) \, \mathrm{d}\nu \, , 
\end{equation}
it has been shown in Appendix~C of Ref.~\onlinecite{daddi2025proc} using the method of residues that this integral admits a closed analytical form, given by
\begin{equation}
    \mathcal{L} = \frac{1}{\sqrt{r^2 + \rho^2 + 2\rho r \cos b + z^2} } \, . 
    \label{eq:L}
\end{equation} 
We define $s^\prime = \mathcal{L}^{-1}(b = \theta + \beta)$, corresponding to the value of~$s$ with $\beta$ replaced by $\pi - \beta$, representing the image position relative to the wall. We then obtain
\begin{equation}
    c_\mathrm{M}^{(1)} = \frac{q_0 R}{s^\prime} \, .
    \label{eq:cM1_PLANAR}
\end{equation}

To obtain the solution for the second reflection, we first require the vector $\bm{C}_\mathrm{M}^{(1)}$ from Eq.~\eqref{eq:CM1_vec}, which simplifies for $\alpha = \pi/2$ to
\begin{equation}
\hspace{-0.1cm}
 \bm{C}_\mathrm{M}^{(1)} = 
 \begin{pmatrix}
    2\nu\sh(2\beta\nu)\sin\beta+\ch(2\beta\nu)\cos\beta \\
    \ch(2\beta\nu)\sin\beta - 2\nu\sh(2\beta\nu) \cos\beta
 \end{pmatrix} \sch(\pi\nu) \, . 
\end{equation}

Thus, $\bm{C}_\mathrm{M}^{(2)}$ from Eq.~\eqref{eq:CM2} can be written as
\begin{equation}
    \bm{C}_\mathrm{M}^{(2)} = \frac{1}{4} \, \cos^2\beta
    \int_0^\infty \bm{C}_\mathrm{M}^{(1)} \, \mathrm{d}\nu \, .
\end{equation}
This integral has a closed analytical form, given by
\begin{equation}
    \int_0^\infty \bm{C}_\mathrm{M}^{(1)} \, \mathrm{d}\nu = 
    \frac{1}{2} \begin{pmatrix}
        \sec^2\beta \\
        0
    \end{pmatrix} .
\end{equation}

Using these results, it follows from Eq.~\eqref{eq:c2M} that the second reflection is given by
\begin{equation}
    c_\mathrm{M}^{(2)} 
    = -\frac{q_0 \epsilon^2}{8} \left( \frac{R}{s} \right)^2 \hat{\bm{s}} \cdot \hat{\bm{x}} \, . 
    \label{eq:cM2_PLANAR}
\end{equation}

Finally, the phoretic velocity is obtained directly from Eq.~\eqref{eq:V_M} as
\begin{equation}
    \bm{V}_\mathrm{M} = \frac{\mu q_0 \epsilon^2}{4R} \, \hat{\bm{x}} ,
    \label{eq:V_M_PLANAR}
\end{equation}
Accordingly, the velocity has a single component in the direction normal to the wall.

\subsection{Source dipole}

By setting $\alpha = \pi/2$ in the expression for the coefficients corresponding to the source dipole in Eqs.~\eqref{eq:coeffs_Dipole}, we obtain
\begin{subequations}
    \begin{align}
    \Lambda_\mathrm{MS}^{(1)} &= \nu \sin\delta  \ch(\nu\beta) \sin(\lambda-\beta) \, , \\
\Lambda_\mathrm{MC}^{(1)} &= \nu \sin\delta  \sh(\nu\beta) \sin(\lambda-\beta) \, ,
\\
    \mathrm{H}_\mathrm{DS}^{(1)} &= \sin\delta  \sh(\nu\beta) \cos(\lambda-\beta) \, , \\
\mathrm{H}_\mathrm{DC}^{(1)} &= \sin\delta \ch(\nu\beta) \cos(\lambda-\beta) \, ,
\\
\Delta_\mathrm{DS}^{(1)} &= \cos\delta \sh(\nu\beta) \, , \\
\Delta_\mathrm{DC}^{(1)} &= \cos\delta \ch(\nu\beta) \, .
\end{align}
\end{subequations}
Then, the three functions defined in Eqs.~\eqref{eq:zeta_xi_phi} take their final form
\begin{subequations}
    \begin{align}
\zeta_\mathrm{D}^{(1)} &= \nu \sin\delta  \sin(\lambda-\beta)\sh\left((\beta+\theta)\nu\right)\, , \\
\xi_\mathrm{D}^{(1)} &= \sin\delta \cos(\lambda-\beta) \ch\left((\beta+\theta)\nu\right)  \, , \\
\phi_\mathrm{D}^{(1)} &= \cos\delta \ch\left((\beta+\theta)\nu\right) \, .
\end{align}
\end{subequations}
This yields the expression for $\psi_\mathrm{D}^{(1)}$ as defined in Eq.~\eqref{eq:psiD}.

From Eq.~\eqref{eq:cD1_sol}, the first reflection of the concentration field can be expressed using the function $\mathcal{L}$ defined in Eq.~\eqref{eq:L} as
\begin{equation}
    c_\mathrm{D}^{(1)} = \left. q_1 R^2 \left( \sin\delta \, \mathcal{D}_\parallel - \cos\delta \, \frac{\partial}{\partial z}  \right) \mathcal{L} \, \right|_{b=\theta+\beta} ,
\end{equation}
where the differential operator acting along the in-plane direction is defined as
\begin{equation}
    \mathcal{D}_\parallel = \sin(\lambda-\beta)\, \frac{1}{\rho} \frac{\partial}{\partial b} + 
    \cos(\lambda-\beta) \, \frac{\partial}{\partial \rho} \, . 
\end{equation}

We obtain, after simplification,
\begin{equation}
    c_\mathrm{D}^{(1)} = q_1 \left( \frac{R}{s^\prime} \right)^2
   \hat{\bm{s}}^\prime \cdot \left( \bm{u}^\parallel - \bm{u}^\perp \right) , 
   \label{eq:cD1_PLANAR}
\end{equation}
with the normal vector $\hat{\bm{s}}^\prime$ defined as $\hat{\bm{s}}$ with $\beta$ replaced by $\pi - \beta$, representing the position vector from the particle's image relative to the wall.
In addition,
\begin{equation}
    \bm{u}^\parallel = \begin{pmatrix}
        0 \\
        \sin\delta \sin\lambda \\
        \cos\delta
    \end{pmatrix}, 
    \quad
     \bm{u}^\perp = \begin{pmatrix}
     \sin\delta\cos\lambda \\
     0 \\
     0 
     \end{pmatrix} ,
\end{equation}
correspond to the parallel and perpendicular projections of the orientation vector relative to the wall.
Note that these vectors are not normalized vectors.

For the calculation of the second reflection, we use the vector $\bm{C}_\mathrm{D}^{(1)}$ from Eq.~\eqref{eq:CD1_vec}, which takes the following form when $\alpha = \pi/2$
\begin{widetext}
    \begin{equation}
        \bm{C}_\mathrm{D}^{(1)} = 
        \begin{pmatrix}
            \left( \left( N_+\cos\lambda-N_-\cos(\lambda-2\beta)\right) \ch(2\beta\nu) - 4\nu \sin(\lambda-2\beta) \sh(2\beta\nu) \right)\sin\delta  \\
            \left( \left( N_+\sin\lambda+N_-\sin(\lambda-2\beta)\right) \ch(2\beta\nu) - 4\nu \cos(\lambda-2\beta) \sh(2\beta\nu) \right)\sin\delta  \\
            \left( 4\nu^2+1 \right) \cos\delta \ch(2\beta\nu)
        \end{pmatrix} \sch(\pi\nu) \, .
    \end{equation}
\end{widetext}

Thus, it follows from Eq.~\eqref{eq:CD2} that
\begin{equation}
    \bm{C}_\mathrm{D}^{(2)} = -\frac{1}{16} \, \cos^3\beta \int_0^\infty \bm{C}_\mathrm{D}^{(1)} \, \mathrm{d}\nu \, .
\end{equation}
The integral in the above equation can be evaluated analytically as
\begin{equation}
    \int_0^\infty \bm{C}_\mathrm{D}^{(1)} \, \mathrm{d}\nu  =
    \left( \bm{u}^\parallel + 2\bm{u}^\perp \right)  \sec^3\beta \, ,
\end{equation}
so that
\begin{equation}
    \bm{C}_\mathrm{D}^{(2)} = -\frac{1}{16} \left( \bm{u}^\parallel + 2\bm{u}^\perp \right) .
    \label{eq:cD2_PLANAR}
\end{equation}

Using these results, the second reflection is obtained from Eq.~\eqref{eq:c2D} as
\begin{equation}
    c_\mathrm{D}^{(2)} = \frac{q_1 \epsilon^3}{16} \left( \frac{R}{s} \right)^2 \left( \bm{u}^\parallel + 2\bm{u}^\perp \right) \cdot \hat{\bm{s}} \, .
\end{equation}

The phoretic velocity is readily obtained from Eq.~\eqref{eq:V_D} as
\begin{equation}
    \bm{V}_\mathrm{D} = -\frac{\mu q_1 \epsilon^3}{8R} \left( \bm{u}^\parallel + 2\bm{u}^\perp \right) .
    \label{eq:V_D_PLANAR}
\end{equation}

These results are fully consistent with the literature, in particular with Ref.~\onlinecite{ibrahim15}, where the first-reflection solutions given by Eqs.~\eqref{eq:cM1_PLANAR} and~\eqref{eq:cD1_PLANAR} correspond to Eq.~(13) of that work, the second-reflection solutions given by Eqs.~\eqref{eq:cM2_PLANAR} and~\eqref{eq:cD2_PLANAR} correspond to Eq.~(14), and the phoretic velocities given by Eqs.~\eqref{eq:V_M_PLANAR} and~\eqref{eq:V_D_PLANAR} correspond to Eq.~(15). We recall that $q_0$ and $q_1$ are obtained from Eq.~\eqref{eq:q_ell}.

\section{Superposition approximation}
\label{sec:superposition}

\begin{figure}
    \centering
    \includegraphics[width=0.9\linewidth]{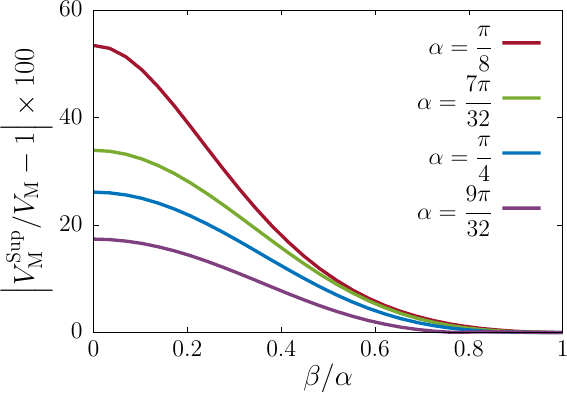}
    \caption{
    Relative error in the magnitude of the phoretic velocity induced by a source monopole, as predicted using the superposition approximation of Eq.~\eqref{eq:VM_Sup}, for various values of~$\alpha$.
    }
    \label{fig:super_M}
\end{figure}

The superposition approximation originally introduced by Oseen~\cite{oseen28} in the context of hydrodynamic mobility provides a framework to obtain approximate expressions for the induced phoretic velocity in the presence of two obstacles. This approach has been widely used in the literature to approximate the mobility between two parallel planar walls~\cite{faxen22,dufresne01,swan10, daddi16b, daddi18jpcm} or between two coaxially positioned rigid disks~\cite{daddi2020axisymmetric,daddi2021steady,daddi2022stokeslet}. In the present context, the idea is to postulate that the velocity of the phoretic particle in wedge geometry can be approximated as the superposition of the effects of each wall, using the results in the planar-wall limit of infinite extent re-derived in Sec.~\ref{sec:planar_boundary}.

\subsection{Source monopole}

The upper wall of the wedge located at $\theta=\alpha$ has normal vector $\hat{\bm{n}}_1=(\sin\alpha,-\cos\alpha,0)$ and the lower wall located at $\theta=-\alpha$ has normal vector $\hat{\bm{n}}_2=(\sin\alpha,\cos\alpha,0)$. It follows from Eq.~\eqref{eq:V_M_PLANAR}, which provides the induced velocity due to monopole interactions near a wall of normal vector $\hat{\bm{x}}$, that the induced velocity obtained using the superposition approximation is given by
\begin{subequations} 
\label{eq:VM_Sup}
    \begin{align}
    {V_\mathrm{M}^\parallel}^\text{Sup} &=
    \frac{\mu q_0 \epsilon^2}{4R}
    \, \chi^2 \sin(2\alpha) \, , \\
    {V_\mathrm{M}^\perp}^\text{Sup} &=
    \frac{\mu q_0 \epsilon^2}{4R}
    \left( 1-\chi^2 \cos(2\alpha) \right) , 
\end{align}
\end{subequations}
where \begin{equation}
    \chi = \frac{\sin(\alpha-\beta)}{\sin(\alpha+\beta)} \, .
\end{equation}
Here, the parallel and normal components relative to the upper wall are defined by Eq.~\eqref{eq:V_M_para_Perp}.

In Fig.~\ref{fig:super_M}, we show the relative error in the magnitude of the monopole contribution to the induced phoretic velocity predicted using the superposition approximation for various values of~$\alpha$. The error is largest in the midplane ($\beta=0$) and decreases monotonically to zero as $\beta$ approaches~$\alpha$, corresponding to the single-wall limit. Within the considered range of opening angles, smaller $\alpha$ results in larger errors, with a maximum relative error exceeding 50\% for $\alpha=\pi/8$ and decreasing to about 18\% for $\alpha=9\pi/32$. Hence, the superposition approximation is generally not recommended for accurately predicting the phoretic velocity due to monopole interactions when $\beta/\alpha$ is small.

\subsection{Source dipole}

\begin{figure}
    \centering
    \includegraphics[width=0.9\linewidth]{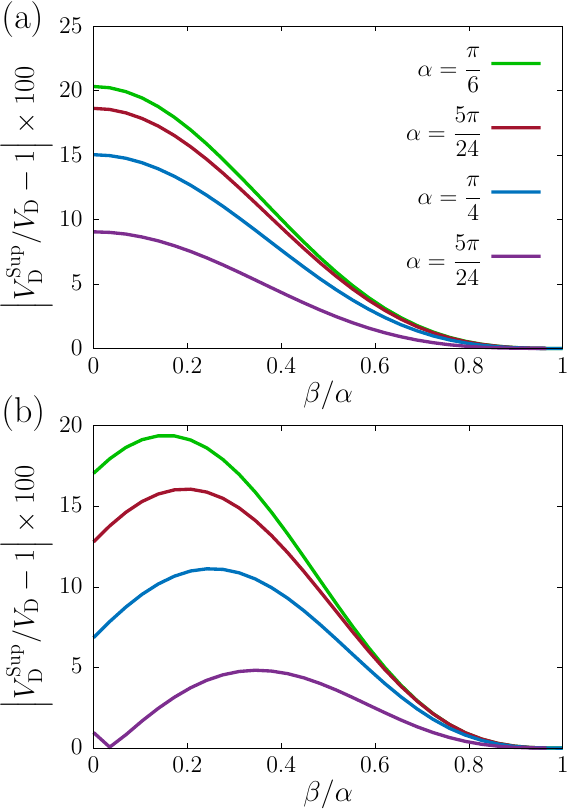}
    \caption{
Relative error in the magnitude of the phoretic velocity induced by a source dipole, predicted using the superposition approximation: (a) for motion along the axial direction with $\delta=0$, as given by Eq.~\eqref{eq:VD_Sup_1}, and (b) for motion in the radial–azimuthal plane with $\delta=\pi/2$ and $\lambda=\alpha$, as given by Eq.~\eqref{eq:VD_Sup_2}.
    }
    \label{fig:super_D}
\end{figure}

For the dipole contribution, the projections of the orientation vector parallel and normal to each wall are required. With respect to the upper wall, we have
\begin{equation}
    \bm{u}_1^\parallel = \begin{pmatrix}
        \sin\delta \cos\alpha \cos(\lambda-\alpha) \\
        \sin\delta\sin\alpha \cos(\lambda-\alpha) \\
        \cos\delta
    \end{pmatrix}
\end{equation}
and
\begin{equation}
    \bm{u}_1^\perp = \begin{pmatrix}
        -\sin\delta \sin\alpha \sin(\lambda-\alpha) \\
        \sin\delta\cos\alpha \sin(\lambda-\alpha) \\
        0
    \end{pmatrix} .
\end{equation}
Note that these are not unit vectors. The corresponding expressions for $\bm{u}_2^\parallel$ and $\bm{u}_1^\perp$ are obtained by replacing $\alpha$ with $-\alpha$.

Using Eq.~\eqref{eq:V_D_PLANAR}, we obtain 
\begin{equation}
    \bm{V}_\mathrm{D}^\text{Sup} 
    = -\frac{\mu q_1\epsilon^3}{8R}
    \left( \bm{u}_1^\parallel+2\bm{u}_1^\perp
    + \chi^3 \left( \bm{u}_2^\parallel+2\bm{u}_2^\perp \right)
    \right) .
\end{equation}
In particular, for $\delta=0$ we obtain
\begin{equation}
     {V_\mathrm{D}^\parallel}^\text{Sup}
     = -\frac{\mu q_1\epsilon^3}{8R}
     \left( 1+\chi^3 \right) \, .
     \label{eq:VD_Sup_1}
\end{equation}
For $\delta=\pi/2$ and $\lambda=\alpha$, we obtain
\begin{subequations}
\label{eq:VD_Sup_2}
    \begin{align}
    {V_\mathrm{M}^\parallel}^\text{Sup} &= 
    -\frac{\mu q_1\epsilon^3}{8R}
    \left( 1+\frac{\chi^3}{2} \left( 3-\cos(4\alpha) \right) \right) , \\
     {V_\mathrm{M}^\perp}^\text{Sup} &= 
    -\frac{\mu q_1\epsilon^3}{16R}\,
    \chi^3 \sin(4\alpha) \, .
\end{align}
\end{subequations}

In Fig.~\ref{fig:super_D} we present the relative error in the magnitude of the phoretic velocity induced by a source dipole, estimated using the superposition approximation: (a) for axial motion with $\delta=0$, according to Eq.~\eqref{eq:VD_Sup_1}, and (b) for motion in the radial–azimuthal plane with $\delta=\pi/2$ and $\lambda=\alpha$, according to Eq.~\eqref{eq:VD_Sup_2}.
Here, the maximum errors remain below roughly 20\% across the entire range of values, indicating that the superposition approximation is appropriate for dipole interactions. As with monopole interactions, smaller angles result in larger errors. In the limit $\beta \to \alpha$, the error vanishes, corresponding to the single-wall limit.

\section{Conclusions}
\label{sec:conclusions}

In summary, we have investigated the self-diffusiophoretic propulsion of a catalytically active spherical particle confined within a wedge-shaped domain using a far-field approach.
Our theoretical developments are conducted using the Fourier-Kontorovich-Lebedev transform. By employing the method of images, we obtained the concentration field up to the second reflection, accounting for both monopole and dipole contributions of the particle’s surface activity. From these results, we derived leading-order expressions for the self-induced phoretic velocity and analyzed its dependence on key geometric parameters, including the wedge opening angle and the particle’s position within the domain.

Our analysis shows that the wedge geometry strongly influences both the magnitude and direction of the particle’s velocity, highlighting the critical role of confinement in guiding active particle motion. 
In this study, we focused on deriving the diffusiophoretic contribution to the particle’s self-propulsion. A complete description of particle dynamics, however, also requires accounting for the hydrodynamic contribution, which arises from higher-order singularities of Stokes flow, primarily the source dipole, force dipole, and force quadrupole~\cite{chwang1975hydromechanics, spagnolie12, lopez2014dynamics, mathijssen2016hydrodynamics, daddi2019frequency, sprenger2020towards}. While expressions for these contributions are available for a planar wall with no-slip boundary conditions~\cite{spagnolie12}, to our knowledge no analogous results exist for wedge-like geometries. A systematic derivation of these higher-order singularities in wedge confinement warrants a dedicated study, given the length and mathematical complexity of such solutions.
Based on previous studies of planar boundaries~\cite{ibrahim15, ibrahim2016walls}, it has been shown that the leading-order hydrodynamic correction to the translational velocity scales as $\epsilon^3$, while the leading-order correction to the rotational velocity scales as $\epsilon^4$. Therefore, a systematic investigation of swimming trajectories in a wedge geometry would require including these hydrodynamic contributions, as they are of the same order as the dipolar contribution to the phoretic velocity.
Moreover, our analysis was limited to the case of constant phoretic velocity, thereby excluding rotational motion induced by phoretic interactions. 
In the case of non-uniform phoretic mobility, the far-field hydrodynamics corresponds to a force-dipole, decaying as $s^{-2}$, which leads to velocity corrections scaling as $\epsilon^2$. Consequently, in this scenario, hydrodynamic effects cannot be neglected, even for the leading-order expressions of the phoretic velocity.
It would be of interest in future work to investigate the effects of mobility moments on particle dynamics and explore the coupling between translational and rotational motion in wedge confinement.
However, it is worth noting that the exact solution derived by Nasouri and Golestanian~\cite{nasouri2020exact, Nasouri2020} for two equal-sized spheres at any separation (including the near-field), which accounts for both hydrodynamics and concentration exactly, shows that removing the hydrodynamic contribution while retaining only the concentration part has little impact on the overall results, introducing only minor quantitative differences.

The results presented in this study provide new insight by showing how the wedge geometry modifies the solute field, which in turn affects the magnitude and direction of the phoretic motion. This analysis clarifies how the wedge opening angle and the particle position within the wedge influence the resulting particle dynamics.
We note that, although the semi-analytical FKL approach does not reduce the computational effort compared to a fully numerical solution, it provides clear physical insight into the role of successive reflections and their contributions to particle dynamics—information that is generally obscured in purely numerical methods. Moreover, it offers a benchmark for validating future numerical or approximate models.

The present analysis primarily focuses on the far-field description of the solute distribution and the resulting phoretic motion. While this framework provides a clear understanding of how confinement by a wedge geometry modifies the solute field through successive reflections at the boundaries, biological environments are often characterized by strong near-field interactions and complex boundary conditions. In such situations, the detailed structure of the solute field close to the particle and the confining surfaces is expected to play a dominant role. Therefore, extending the present approach to account for near-field effects and more realistic boundary conditions represents an important direction for future work in order to better assess the relevance of these mechanisms in biological media.
A full assessment of the accuracy of the far-field approach used in this work would require exact analytical treatments or fully resolved numerical simulations, prior experience with planar and spherical interfaces indicates that this approximation can yield remarkably accurate predictions of hydrodynamic mobilities even at distances comparable to the particle diameter~\cite{daddi16, daddi2017hydrodynamic_asy, daddi2018creeping}. This discussion highlights avenues for future work, including the use of exact analytical frameworks such as bipolar coordinates.

An example experimental setup to test the phoretic motion near a wedge could use spherical photo-catalytic particles with uniform phoretic mobility and a partially illuminated surface to control activity. The active colloid would be suspended in a viscous, transparent fluid, such as water. A microscale wedge made of PDMS or glass, with adjustable opening angles, would be placed inside a microfluidic chamber or microsized tank to provide confinement. The motion of the active colloid could be directly observed and recorded using a high-resolution microscope with a camera positioned to capture the plane of interest. Control experiments with passive colloids of the same size could be performed to isolate phoretic effects from background fluid motion.

This study provides a systematic
framework for calculating the contributions to the phoretic velocity arising from concentration disturbances near corners, providing a perspective relevant to the design of microfluidic devices and the control of active particles in complex geometries. Future work could extend this approach to account for hydrodynamic interactions, finite-size effects, and more complex boundary shapes.

\section*{Author contribution}
A.D.M.I.\ and R.G.\ designed the research.
A.D.M.I.\ performed the analytical calculations, prepared the figures, and wrote the paper.
A.D.M.I.\ and R.G.\ reviewed, edited, and contributed to the refinement of the manuscript.

\begin{acknowledgments}
R.G.\ acknowledges support from 
the Max Planck Center Twente for Complex Fluid Dynamics, the Max Planck School Matter to Life, and the MaxSynBio Consortium, all of which are jointly funded by the Federal Ministry of Education and Research (BMBF) of Germany and the Max Planck Society.
\end{acknowledgments}

\section*{Declaration of interests}
The authors report no conflict of interest.

\vspace{1cm}
\section*{Data availability}
The data that support the findings of this article are not publicly available. The data are available from the authors upon reasonable request.

\section*{Author ORCIDs}
\begin{itemize}
    \item Abdallah Daddi-Moussa-Ider: \href{https://orcid.org/0000-0002-1281-9836}{0000-0002-1281-9836}
    \item Ramin Golestanian: \href{https://orcid.org/0000-0002-3149-4002}{0000-0002-3149-4002}
\end{itemize}

\appendix*

\section{Analytical Evaluation of Improper Integrals}

\subsection{Evaluation of the integral in Eq.~\eqref{eq:f_M_integral}}

We consider the improper integral
\begin{equation}
    f_1(\alpha) = \int_0^\infty 
    \left( \cth(\alpha\nu) \th(\pi\nu) - 1 \right)
    \mathrm{d}\nu \, .
    \label{eq:app:int1}
\end{equation}
We rewrite this integral in the form
\begin{equation}
    f_1(\alpha) = \int_0^\infty
    \sh( \nu(\pi -\alpha)) \sch(\pi\nu)\csh(\alpha\nu) \,
    \mathrm{d}\nu \, .
\end{equation}
By using the series representations 
\begin{align}
    \sch(\pi\nu) &= 2\sum_{n=0}^\infty (-1)^n e^{-(2n+1)\pi\nu} \, , \\
    \csh(\alpha\nu) &= 2 \sum_{m=0}^\infty  e^{-(2m+1)\alpha\nu} \, , 
\end{align}
together with the integral
\begin{equation}
    \int_0^\infty e^{-a\nu} \sh(b\nu)\, \mathrm{d}\nu= \frac{1}{2}
    \left( \frac{1}{a-b} - \frac{1}{a+b} \right) ,
    \label{eq:int_sh_exp}
\end{equation}
where $0\le b < a$, we rewrite the integral as a double series of the form
\begin{equation}
    f_1(\alpha) = \sum_{m,n} (-1)^n 
    \left( \frac{1}{A_{mn}} - \frac{1}{B_{mn}} \right) , 
\end{equation}
where $A_{mn}=\pi n+\alpha m+\alpha$ and $B_{mn} = \pi n+ \alpha m + \pi$, as defined in the main body of the paper.

We now evaluate the integral in the special case of a commensurate semi-opening angle $\alpha_q=\pi/q$.
Since the integrand in Eq.~\eqref{eq:app:int1} is even in $\nu$, the integral from~$0$ to~$\infty$ equals half of the integral over the entire real axis.
We rewrite the integral as
\begin{equation}
    f_1(\alpha_q) = \frac{1}{2} \int_{-\infty}^\infty 
    \left( \cth(\alpha_q\nu) \th(\pi\nu) - 1 \right)
    \mathrm{d}\nu \, ,
\end{equation}
and use the change of variable $u = e^{\frac{2\pi x}q}$ to obtain
\begin{equation}
f_1(\alpha_q)
= \frac {q}{2\pi}\int_0^\infty\frac{u^{n-1}-1}{(u-1)(u^n+1)}\, \mathrm{d}u \, .
\end{equation}
This can be written as a finite series as
\begin{equation}
    f_1(\alpha_q) = \frac {q}{2\pi}\sum_{k=1}^{q-1} \int_0^\infty \frac{u^{k-1}}{u^n+1}\, \mathrm{d}u \, .
\end{equation}
Using the integral
\begin{equation}
    \int_0^\infty\frac{u^{k-1}}{u^q+1}\, \mathrm{d}u = 
    \frac \pi{q} \, \csc \frac{k\pi}q \, , 
    \label{eq:csc_int}
\end{equation}
it follows that 
\begin{equation}
    f_1(\alpha_q) = \frac{1}{2} \, \sum_{k=1}^{q-1} \csc \frac{k \pi}q  \, .
\end{equation}

\subsection{Evaluation of the integral in Eq.~\eqref{eq:f_D_integral}}

We next evaluate the improper integral
\begin{equation}
    f_2(\alpha) = 
   \int_0^\infty
    \left( 4\nu^2+1 \right)
    \left( \cth(\alpha\nu) \th(\pi\nu) - 1 \right)
    \mathrm{d}\nu \, .
    \label{eq:app:int2}
\end{equation}
We follow the same approach by expanding $\sch(\pi\nu)$ and $\csh(\alpha\nu)$ into infinite series and evaluating the resulting integral.
Here, we use the integral in Eq.~\eqref{eq:int_sh_exp} together with
\begin{equation}
    \int_0^\infty \nu^2 e^{-a\nu} \sh(b\nu)\, \mathrm{d}\nu= 
    \frac{1}{(a-b)^3} - \frac{1}{(a+b)^3} \, ,
\end{equation}
where $0\le b < a$, to obtain
\begin{equation}
\hspace{-0.3cm}
    f_2(\alpha) = \sum_{mn} (-1)^n
    \left(  \frac{1}{A_{mn}} - \frac{1}{B_{mn}}
    + \frac{2}{A_{mn}^3}-\frac{2}{B_{mn}^3}
    \right) .
\end{equation}

For a commensurate angle, we adopt the same approach used for $f_1(\alpha_q)$ and rewrite the integral in the form
\begin{equation}
    f_2(\alpha_q) = 
    \frac{q}{2\pi} \int_0^\infty \frac{u^{q-1}-1}{(u-1)(u^q+1)} 
    \left( 1+\frac{q^2}{\pi^2}\, \ln^2 u \right) \mathrm{d}u\, ,
\end{equation}
which may be written, by invoking a finite series, as
\begin{equation}
    f_2(\alpha_q) = \frac{q}{2\pi} \sum_{k=1}^{q-1} \int_0^\infty
    \frac{u^{k-1}}{u^q+1}
    \left( 1 + \frac{q^2}{\pi^2}\, \ln^2 u  \right) \mathrm{d}u\, .
\end{equation}
Using Eq.~\eqref{eq:csc_int} together with
\begin{equation}
    \int_0^\infty \frac{u^{k-1}\ln^2 u}{u^q+1} \, \mathrm{d}u
    =  
    \frac{\pi^3}{q^3} \, \csc \frac{k\pi}{q}
    \left( 2\csc^2 \frac{k\pi}{q} -1\right),
\end{equation}
we finally obtain
\begin{equation}
    f_2(\alpha_q) = \sum_{k=1}^{q-1} \csc^3 \frac{k\pi}{q} \, .
\end{equation}

\subsection{Evaluation of the integral in Eq.~\eqref{eq:g_D_integral}}

This integral 
\begin{equation}
     f_3(\alpha) = \int_0^\infty 
    \left(4\nu^2+3 \right)
    \left( \cth(\alpha\nu) \th(\pi\nu) - 1 \right)
    \mathrm{d}\nu \, ,
\end{equation}
is closely related to the two integrals defined by Eqs.~\eqref{eq:app:int1} and \eqref{eq:app:int2} evaluated previously and can be expressed as
\begin{equation}
    f_3(\alpha) = f_2(\alpha) + 2f_1(\alpha) \, .
\end{equation}

\subsection{Evaluation of the integral in Eq.~\eqref{eq:h_D_integral}}

Finally, we examine the improper integral of the form
\begin{equation}
    f_4(\alpha) = 
    \int_0^\infty
    \nu^2 \left( 1-\th(\alpha\nu)\th(\pi\nu) \right) \mathrm{d}\nu \, .
    \label{eq:app:int4}
\end{equation}
To evaluate this integral, we use the identity
\begin{align}
1 - \th(\pi \nu)\th(\alpha \nu) &= \varphi(\pi\nu)+\varphi(\alpha\nu)-\varphi(\pi\nu)\varphi(\alpha\nu) \, , 
\end{align}
where we have defined
\begin{equation}
    \varphi(x) = \frac{2}{1+e^{2x}} = 2\, \sum_{n=0}^\infty (-1)^n e^{-2(n+1)x} \, , 
\end{equation}
for $x>0$.
Using the integral
\begin{equation}
    \int_0^\infty \nu^2 e^{-a\nu} \, \mathrm{d}\nu = \frac{2}{a^3} \, , 
\end{equation}
for $a>0$, the integral in Eq.~\eqref{eq:app:int4} can be written as
\begin{equation}
    f_4(\alpha) = \frac{1}{2} \left( \frac{1}{\pi^3}+\frac{1}{\alpha^3} \right)
    \sum_{n=0}^\infty \frac{(-1)^n}{(n+1)^3}
    +\sum_{m,n} {h_\mathrm{D}}_{mn}
    \, ,
\end{equation}
where
\begin{equation}
     {h_\mathrm{D}}_{mn} = 
    \frac{(-1)^{n+m+1}}{\left( \pi n+\alpha m+\pi+\alpha \right)^3} \, .
\end{equation}
This can be further simplified by invoking the Riemann zeta function, yielding
\begin{align}
    f_4(\alpha) = \frac{3}{8}\, \zeta(3) \left( \frac{1}{\pi^3} + \frac{1}{\alpha^3} \right) 
    +\sum_{m,n} {h_\mathrm{D}}_{mn} \, .
\end{align}

When considering a commensurate semi-opening angle $\alpha_q=\pi/q$, the analysis becomes more involved than in the cases discussed previously. The final result depends on the parity of $q$.

The key idea is to apply the method of residues to evaluate the integral in the complex plane. We therefore consider the function of a complex variable
\begin{equation}
    g(z) = \frac{1}{2} \, e^{a z} 
    \left( 1-\th(\pi z) \th \left( \frac{\pi z}{k} \right) \right) .
    \label{eq:g_complex_int}
\end{equation}
We define the integral
\begin{equation}
    G(\alpha_q, a) = \int_{-\infty}^\infty g(x)\, \mathrm{d}x \, .
\end{equation}
It can clearly be seen that
\begin{equation}
    f_4(\alpha_q) = 
    \left. \frac{\partial^2 G(\alpha_q, a)}{\partial a^2} \right|_{a=0} \, .
\end{equation}

We integrate $g(z)$, defined in Eq.~\eqref{eq:g_complex_int}, along a rectangular contour oriented anticlockwise with vertices at $-R$, $R$, $R+iq$, and $-R+iq$, where $R>0$ is taken to infinity. We denote by $\mathcal{C}_i$, $i\in\{1,2,3,4\}$, the line segments composing the contour, with $\mathcal{C}_1$ corresponding to the integral along the real axis. It is straightforward to show that the integrals along the two vertical segments vanish as $R\to\infty$. We are therefore left with the two horizontal contributions. Owing to the periodicity of $\th$, the integral along the upper horizontal segment $\mathcal{C}_3$ satisfies
\begin{equation}
\int_{\mathcal{C}_3} g(z)\, \mathrm{d}z = e^{iaq}\int_{\mathcal{C}_1} g(z)\, \mathrm{d}z \, .
\end{equation}
The residue theorem then yields
\begin{equation}
    \left( 1-e^{iaq} \right) G(\alpha_q) = 2i\pi \sum \operatorname{Res} \{ g(z) \} \, .
\end{equation}

We evaluate the residues according to the parity of $q$.
For even $q$, there are $q$ simple poles inside the contour arising from $\th(\pi z)$, located at $z = i(k-1/2)$ for $k=1,\dots,q$.
For odd $q$, there are $q-1$ simple poles at $z = i(k-1/2)$ for $k=1,\dots,q$, excluding $k=(q+1)/2$, and a single double pole at $z = iq/2$.
The residues can be evaluated straightforwardly, and by exploiting the symmetry of the hyperbolic tangent, the sum can be truncated at $k=q/2$ for even $q$ and at $k=(q-1)/2$ for odd $q$.
We obtain
\begin{equation}
    G(\alpha_q) = 
    \frac{1}{2\pi} \cfrac{aq}{\sin \cfrac{aq}{2}}
    \left( 1-(-1)^q\right)
    + \sum_{k=1}^{\lfloor q/2\rfloor}  \Gamma_{qk} \, ,
    \label{eq:G}
\end{equation}
where
\begin{equation}
    \Gamma_{qk} = \cfrac{ \sin \left( \cfrac{a}{2} \left( q+1-2k \right) \right) }{\sin \cfrac{aq}{2}} \, \tan \cfrac{(2k-1)\pi}{2q} \, .
\end{equation}
The first term in Eq.~\eqref{eq:G} arises from the residue at $z=iq/2$ and vanishes when $q$ is even. Taking the second derivative of Eq.~\eqref{eq:G} with respect to $a$ and then letting $a\to 0$ yields the integral $f_4(\alpha_q)$.

By taking higher-order derivatives of $G$, the approach outlined here also permits the evaluation of integrals of the form
\begin{equation}
    \int_0^\infty
    \nu^{2n} \left( 1-\th(\alpha_q\nu)\th(\pi\nu) \right) \mathrm{d}\nu = \frac{\partial^{2n}G(\alpha_q)}{\partial a^{2n}} \, , 
\end{equation}
for $n\ge 0$.

\bibliography{biblio}

\end{document}